\documentclass[psfig,a4paper]{article}
\usepackage{graphicx}
\usepackage{amsmath,latexsym}
\usepackage{psfig}

\def\proof{{\bf{Proof }}} 
\def\fin{\hspace{\stretch{1}}$\Box$}

\newcommand{\be}{\begin{equation}}
\newcommand{\eq}[1]{\begin{equation} #1 \end{equation}}
\newcommand{\ee}{\end{equation}}

\def\Tr{{\rm Tr}}

\def\bin{\{0,1\}}
\def\base{\{+,\times\}}
\def\a{\vec a}
\def\b{\vec b}
\def\g{\vec g}
\def\h{\vec h}
 \def\P{{\mathcal{P}}}
\def\k{\vec\kappa}
\def\B{{\mathcal{B}}}
\def\H{{\mathcal{H}}}
\def\F{{\mathcal{F}}}
\def\Id{\textbf{1}}
\def\G{{\mathcal{G}}}
\def\C{{\mathcal{C}}}
\def\x{\vec x}
\def\s{\vec s}
\def\y{\vec y}
\def\S{{\mathcal{S}}}

\def\CS{C_{\s,\g(\overline{E}\cup M)}}
\def\CSK{C_{\s,\k,\g(\overline{E}\cup M)}}

\def\gd{g(\delta,\tau_f,p_R,n)}

\def\rhos{\tilde{\rho}_{\s,\g(\overline{E}\cup M)}}
\def\rhosk{\tilde{\rho}_{\s,\k,\g(\overline{E}\cup M)}}

\def\M{{\mathcal{M}}}
\def\Z{{\mathcal{Z}}}

\def\K{{\mathcal{K}}}
\def\L{{\mathcal{L}}}
\def\R{{\mathcal{R}}}
\def\X{{\mathcal{X}}}
\def\Y{{\mathcal{Y}}}
\def\da{d_1}
\def\db{d_2}
\def\khi{\chi}
\def\phiv{\tilde{\phi}_v}
\def\Vect{\textrm{Span}}
\def\I{{\mathcal{I}}}

\newcommand{\piz}[1]{\Pi_0(#1)}

\newcommand{\bpio}[1]{\overline{\Pi}_1(#1)}
\newcommand{\pio}[1]{{\Pi}_1(#1)}
\newcommand{\piw}[1]{\Pi_w(#1)}
\newcommand{\bpiw}[1]{\overline{\Pi}_w(#1)}
\newcommand{\rv}[1]{\boldsymbol{#1}}
\newcommand{\pr}[1]{{\rm P}_{\rv{#1}}}
\newcommand{\pc}[2]{{\rm P}_{\rv{#1}\,|\,#2}}
\def\D{\mathcal{D}}

\def\A{\mathcal{A}}

\def\st{\,:\,}
\def\W{{\mathcal{W}}}
\def\tb{\tilde{\b}}

\def\th{\h}
\def\alt{}

\def\hd{h(\delta,\tau_f,p_R,n)}

\def\mp{modified protocol\,}
\def\Nat{{\bf I \!N}}
\def\Real{{\bf R}}

\def\l{l}
\def\lh{\widehat{l}}
\def\ha{\widehat}
\def\tauh{\hat{\tau}}
\def\dwh{\widehat{d}_w}
\def\ch{\check}
\def\Gh{\ha{\G}}
\def\Ch{\ha{\C}}

\def\M{{\mathcal{M}}}
\def\Mc{\rv{\M}=\M}
\def\CM{C_{\M}}
\def\Pmpr{\pr{\M P R}(\M,P,R)}

\def\Mmax{M_{max}}

\newcommand{\state}[1]{\left |\, \Psi (#1)\right \rangle}

\newcommand{\ket}[1]{\big | \, #1 \big \rangle}
\newcommand{\bra}[1]{\big \langle #1 \, \big |}
\newcommand{\proj}[1]{\ket{#1}\bra{#1}}

\newtheorem{definition}{Definition}
\newtheorem{theorem}{Theorem}
\newtheorem{property}{Property}
\newtheorem{lemma}{Lemma}

\def\share{\mbox{\bf share}}
\def\valid{\mbox{\bf valid}}
\def\U{U^{\a(E\cap\overline{M})}}

\def\rmin{r_{min}}
\def\Q{{\mathcal{Q}}}

\begin{document}

\title{Unconditional Security of Practical Quantum Key Distribution}
 \author{Hitoshi Inamori\footnote{ Centre for Quantum Computation,
Clarendon Laboratory, Oxford, United Kingdom}, Norbert
L\"utkenhaus\footnote{ Helsinki Institute of Physics, Helsinki,
Finland \newline\indent present affiliation: MagiQ Technologies, Inc.,
New York NY, United States} and Dominic Mayers\footnote{ NEC Research
Institute, Princeton NJ, United States\newline\indent Computer Science
Department, Maharishi University of Management, USA}}
\date{\today}
\maketitle

%%%%%%%%%%%%%%%%%%%%%%%%%%%%%%%%%%%%%%%%%%%%%%%%%%%%%%%%%%%%%%%%%%
%	SECTION 1	Introduction
%%%%%%%%%%%%%%%%%%%%%%%%%%%%%%%%%%%%%%%%%%%%%%%%%%%%%%%%%%%%%%%%%%

\section{Introduction}

We present a proof of unconditional security of a practical quantum
key distribution protocol. It is an extension of a previous result
obtained by Mayers~\cite{mayers96a,mayers98a}, which proves
unconditional security provided that a perfect single photon source is
used. In present days, perfect single photon sources are not available
and, therefore, practical implementations use either dim laser pulses
or post-selected states from parametric downconversion. Both practical
signal types contain multi-photon contributions which characterise the
deviation from the ideal single-photon state. This compromise
threatens seriously the security of quantum key distributions when the
loss rate in the quantum channel is
high~\cite{huttner95a,yuen96a,brassard00a}. Security of such practical
realisation has nevertheless been proven in \cite{nl00a} against
restricted type of eavesdropping attacks. The salient idea used in
\cite{nl00a} is that data associated with multiple photon signals are
revealed to a possible eavesdropper, without the legitimate user's
knowledge. We show here that this model can be combined with Mayers'
proof. The resulting extension guarantees unconditional security of a
realistic quantum key distribution protocol against an enemy with
unlimited classical or quantum computational power.

By now, Mayers' proof has been followed up by other proof of the
security of ideal single-photon quantum key distribution
\cite{biham99suba,shor00a}.  Security assuming some restrictions on
eavesdropper's attack can be found in~\cite{incoherent, ncoherent,
collective}. Security of protocols in which honest participants use
trusted quantum computers can be found in~\cite{lo99a}.

Unconditional security of a protocol means a security against a
cheater with unlimited computational power, quantum or classical. In
other words, it means that there is no condition on the cheater.  It
does not mean that there is no condition on the apparatus used by the
honest participants.  This last interpretation would be equivalent to
say that we know nothing about the protocol that is actually
implemented.  So, each proof of unconditional security must use a
different type of assumptions on these apparatus.  Mayers' original
proof applies to an unrestricted eavesdropper's attack on the quantum
signals, but assumes the source used in the protocol is perfect. In
particular, it assumes that the source emits single photon pulses. In
this paper, we present a derivation of the proof in which the last
assumption is relaxed: we still consider sources that perform perfect
polarisation encoding, but each signal carries now a random number of
photons in the ideal polarisation mode. The random variables giving
the numbers of photons in the pulses are assumed to be identically and
independently distributed, and we require that an upper-bound on the
probability that a pulse contains several photons is known. As in
Mayers' original paper, there is no assumption on the quantum channel
nor on the detection unit, except that, given an input quantum state
of any signal, the detector's probability of detecting a signal does
not depend on the choice of the measurement basis.  A more detailed
discussion about assumptions in quantum cryptography together with a
new approach to this problem, especially the problem of an untrusted
BB84 source, can be found in \cite{MY97}.

This paper is divided into two parts. In the first part we define the
assumptions of our proof, the protocol we refer to and the security
notion. We then give the result of our proof which give the precise
quantitative meaning of our security proof together. In that step the
necessary parameters of the protocol leading to secure quantum key
cryptography are given.  We illustrate the results by giving the
asymptotic formulas for the limit of long keys which show, how many
secure bits the protocol will obtain for a given error rate of an
experimental set-up as a function of the source parameters and the
error rate. In the second part, we present the detailed proof of the
statements of the first part. We have chosen to give all details of
this proof to make it self contained, although it follows closely
Mayers original work.  The readers are invited to refer to the
original paper~\cite{mayers98a} where a simpler situation was
analysed, to get an insight into the main idea of the proof.

%%%%%%%%%%%%%%%%%%%%%%%%%%%%%%%%%%%%%%%%%%%%%%%%%%%%%%%%%%%%%%%%%%
%	SECTION 2 	Security in Quantum Key Distribution
%%%%%%%%%%%%%%%%%%%%%%%%%%%%%%%%%%%%%%%%%%%%%%%%%%%%%%%%%%%%%%%%%%

\section{Security in Quantum Key Distribution}

The r\^ole of key distribution between two distant legitimate parties,
traditionally called Alice and Bob, is to generate a shared random
binary string, called the private key, that is guaranteed to be known
only by the legitimate parties. A non-authorised party, traditionally
called Eve, should not be able to obtain any information about the
private key. More precisely, for any eavesdropping strategy Eve
chooses, the conditional entropy of the private key, given the data
Eve acquires during the protocol, should be very close to the maximum
entropy, corresponding to a uniformly and independently distributed
key. One requirement for this is that the conditional probability of
the private key given Eve's data must be very close to the uniform
distribution.  Note that it is not sufficient to impose that the
private key be independent of the data Eve acquires: a key
distribution protocol that returns a specific value for the private
key with high probability does not provide any privacy, even if Eve is
inactive during the key distribution.

Quantum key distribution protocols do not allow Alice and Bob to share
a private key in all circumstances. For example, Eve can usually block
signals between the two parties. But even if the signals arrive, Alice
and Bob cannot always create a secure key using them. As shown in
\cite{brassard00a}, it is in principle not possible to create a secure
key with the BB84 protocol (using ideal signals) once the error rate
exceeds 25 \%. This is true for any post-processing of the data in the
sense of advantage distillation or similar ideas.  It is therefore
characteristic for any full protocol (including the classical
post-processing of the data) that it can deliver a secure private key
only as long as the parameters describing the transmission of the
quantum channel (like the error rate) are within a certain parameter
region.

Any protocol therefore provides a validation test that tells whether a
key can be generated with unconditional privacy. A key is created only
if the test is passed. Otherwise the session is abandoned. Naturally,
one would like to find an entropic bound given the validation test is
passed. However, it is known that such a bound is inappropriate for
the protocol we consider in this paper (see for example
\cite{biham99suba}): there are simple attacks that give full knowledge
about the private key, although with very small probability of
success. It is therefore important to choose a good measure of privacy
which nevertheless reflects our basic intuition.

We follow Mayers' proof and define formally a key even in the cases
that the validation test is not passed. For this purpose Bob formally
chooses with uniform distribution a binary sequence as key whenever
the test fails. We then bound Eve's entropy on this always defined
key, conditioned on her knowledge, to be arbitrarily close to the
maximal value. Naturally, in that case Alice and Bob do not share a
key, but this is unimportant since they are aware of it.

This choice of security notion assures that Eve's conditional entropy
is close to the maximal amount, but this situation can arise from two
different scenarios: either Eve applies only gentle eavesdropping,
which passes the validation tests and gives her basically no
information, or she applies massive eavesdropping, which basically all
the time fails the validation test, but in the unlikely event of
passing the test, it might reveal substantial amount of
information. Nevertheless, in both cases the key will be safe, since
in the first scenario Eve has no information on the key, while in the
second case, the probability of success will be, in a quantified way,
extremely low.

Another important aspect of security of quantum key distribution
protocols is the integrity or the faithfulness of the distributed
key. We must require that whatever Eve does, it is very unlikely that
Alice and Bob fail to share an identical private key while the
validation test is passed. One way this situation might arise is the
error correction procedure (which is a typical ingredient of a full
protocol) failing to correct all errors, for example because of an
unusual error distribution.

Finally, we consider families of protocols for which a parameter,
quantifying the amount of a resource used in a protocol, characterises
its security. Usually, the higher this {\em security parameter's}
value is, the higher is the level of security, but also the amount of
a resource required by the protocol. In the protocol we consider the
number of quantum signals sent by Alice as security parameter.

 We now give a formal definition of security. For this we will
introduce some notation. A random variable will always be denoted by a
bold letter, and values taken by this random variable by the
corresponding plain letter. Only discrete random variables will be
considered in this paper. The probability distribution of a random
variable $\rv{x}$ is denoted by $\pr{x}$,
i.e. $\pr{x}(x)=\Pr(\rv{x}=x)$ is the probability that $\rv{x}$ takes
the value $x$. The joint distribution of two random variables $\rv{x}$
and $\rv{y}$ is denoted by $\pr{ x y}$, i.e. $\pr{x
y}(x,y)=\Pr(\rv{x}=x,\rv{y}=y)$. The conditional probability of
$\rv{x}$ given an event $\mathcal{E}$ with positive probability is
denoted by $\pc{x}{\mathcal{E}}$, i.e. $\pc{x}{{\mathcal{E}}}(x) =
\Pr(\rv{x}=x |{\mathcal{E}})$. The conditional probability of $\rv{x}$
given that $\rv{y}$ takes a value $y$ is denoted by $\pc{x}{\rv{y}=y}$
whenever $\pr{y}(y)>0$, i.e. $\pc{x}{\rv{y}=y}(x) =
\Pr(\rv{x}=x|\rv{y}=y)=\frac{\pr{x y}(x,y)}{\pr{y}(y)}$, whenever
$\pr{y}(y)$ is positive. Let $f$ be a function defined on the image of
$\rv{x}$. When no confusion is possible, the notation $\rv{f}$ will be
adopted to denote the random variable $f(\rv{x})$.

We will denote by $\rv{\k}$ the random variable giving the private key
 generated in a key distribution session. The key is a string of $m$
 bits where $m$ is a positive integer specified by the legitimate
 users. That is $\rv{\k}$ takes value in $\bin^m$. We denote by
 $\rv{\valid}$ the random variable giving the outcome of the
 validation test and by $\rv{\share}$ the random variable telling
 whether Alice and Bob share an identical private key.  Given an
 eavesdropping strategy chosen by Eve, we denote by $\rv{v}$ the
 random variable giving collectively all data Eve gets during this key
 distribution session. Henceforth, given the eavesdropping strategy
 adopted by Eve, $\rv{v}$ is called the {\em view} of Eve, and we will
 denote by $\Z$ the set of all values $\rv{v}$ may take.

We adopt the following definition of security for quantum key
distribution protocols.

\begin{definition}\label{Definition}
Consider a quantum key distribution protocol returning a key
$\rv{\k}\in\bin^m$ regardless of the outcome of the validation test,
where the length of the key, $m$, is fixed and chosen by the user. We
say that the protocol has \emph{(asymptotic) perfect security} if and
only if:
\begin{itemize}
\item the protocol is parametrised by a parameter $N$ taking value in
$\Nat$ called the \emph{security parameter}, and
\item there exists two functions $\epsilon_1,\, \epsilon_2\,:\,\Nat
\times\Nat \rightarrow \Real^+$ such that $\epsilon_1(N,m)$ and
$\epsilon_2(N,m)$ are vanishing exponentially as $N$ grows (i.e. there
exist $\alpha>0$, $\beta>0$, $N_{min}\in \Nat$ and a function
$f:\,\Nat\rightarrow \Real^+$ such that $\forall N>N_{min},\,
\epsilon_1(N,m),\, \epsilon_2(N,m) < e^{-\alpha N^\beta} f(m)$) , and
\item there exists a function $N_0\;:\;\Nat \rightarrow \Nat$ such
that, for any strategy adopted by Eve,
\begin{eqnarray}
\lefteqn{\forall m, \forall N\geq N_0(m), }\nonumber\\ &\mbox{\bf
(privacy)}& \quad H(\rv{\k} | \rv{v}) \geq m -\epsilon_1(N,m)\\
&\mbox{\bf (integrity)}& \quad \Pr(\neg\rv{\share} \mbox{ \rm and }
\rv{\valid}) \leq \epsilon_2(N,m)
\end{eqnarray}
where $\rv{v}$ is Eve's view given her strategy, and $H(\rv{\k} |
\rv{v})\stackrel{Def}{=}-\sum_{\k,v \,|\, \pr{\k v}(\k,v) >0}\pr{\k
v}(\k,v)\log_2 \pc{\k}{\rv{v}=v}(\k)$ is the Shannon
entropy~\cite{Sha,Wel,Sti} of the key $\rv{\k}$ given Eve's view
$\rv{v}$.
\end{itemize}
\end{definition}
We will show that the protocol presented in the next section will be
secure according to this definition. In particular, this means, that
the protocol creates a key of length $m$ out of $N$ signals. Then, by
choosing $N$ large enough for fixed values of $m$, we can always
assure that Eve's conditional entropy is arbitrarily close to the
maximum amount (privacy). Additionally, with a probability arbitrarily
close to unity, Alice and Bob share the key given that the validation
test is passed (integrity).

%%%%%%%%%%%%%%%%%%%%%%%%%%%%%%%%%%%%%%%%%%%%%%%%%%%%%%%%%%%%%%%%%%%%%%%%%%%%
%	SECTION 3	The Protocol
%%%%%%%%%%%%%%%%%%%%%%%%%%%%%%%%%%%%%%%%%%%%%%%%%%%%%%%%%%%%%%%%%%%%%%%%%%%%

\section{The protocol}
In this section, the quantum key protocol considered in this paper is
described. It is an adaptation of the BB84~\cite{BB84} protocol which
takes into account the usage of an imperfect photon source. Note that
the usage of imperfect source has been discussed as early as the first
experimental implementation of BB84~\cite{bennett92a} in the framework
of restricted types of eavesdropping attacks. We first make precise
which assumptions on the quantum channel we adopt in this paper. Then
we give a formal description of the protocol.

%%%%%%%%%%%%%%%%%%%%%%%%%%%%%%%%%%%%%
%	3.1	Required Technology
%%%%%%%%%%%%%%%%%%%%%%%%%%%%%%%%%%%%%

\subsection{Required technology}
In the original proof~\cite{mayers98a}, Mayers considered a practical
realisation of quantum key distribution prone to noise and signal
loss. However, the legitimate parties were assumed to be using a
perfect single photon source -- a source that emits exactly one photon
in the chosen polarisation state. No restriction was imposed on the
photo-detection unit used in the protocol, except that given an
incoming signal, the probability of detection was required to be
independent of the basis used to measure the signal. It was argued
in~\cite{mayers98a} that Eve can take advantage of a detection unit in
which the probability of detection depends on the basis chosen for the
measurement, and we will adopt in this paper the same restriction
regarding the detection unit.

The new feature in this paper is that we allow the use of imperfect
source of photons in the following sense: given a polarisation state
specified by the user, the source emits photons exactly in the
specified polarisation state, but in a mixture of Fock states. That
is, the source emits $n$ photons in the given polarisation state with
probability $p_n$, where $n\in\Nat$ and $p_0, p_1, p_2,\ldots$ is a
probability distribution. The user does not have to know how many
photons were actually emitted. The only restriction we impose is that
an upper bound $\Mmax$ on the number of emitted signals containing
several photons is known within a confidence limit given by the
(small) probability $Pr(M>\Mmax)$.  We restrict ourselves to provide
this bound for signals with identically and independently distributed
multi-photon probability $p_M$. In that case we can choose $\Mmax =
(p_M + \tau_M) N$ and obtain $Pr(M>\Mmax)< \exp(-\tau_M^2 N)$, as
explained below. Other methods for providing $\Mmax$ and $Pr(M>\Mmax)$
can be used, where the corresponding terms replace the here derived
and easily identifiable expressions in the subsequent results.

The authors believe this relaxation of requirement has practical
importance, since single photon sources are not yet available, due to
technological limitations. Furthermore, it has been pointed
out~\cite{brassard00a} that in most experimental implementations of
quantum key distribution, the quantum signals transmitted by the
legitimate parties can be described as mixtures of Fock states.

As an example, consider a practical source emitting a \emph{coherent
state} of light in a given polarisation: 
\be \ket{\alpha} =
e^{-\frac{|\alpha|^2}{2}}\sum_{j=0}^{\infty}\frac{\alpha^j}{\sqrt{j!}}\ket{j}\label{coh}
\end{equation}
where $\ket{j}$, $j\in\Nat$ is the number state -- or Fock state --
describing a state of $j$ photons in the considered polarisation
(Therefore, for $\alpha\neq 0$, a coherent state has an indefinite
number of photons). If we write $\alpha =|\alpha|e^{i\phi}$,
$|\alpha|$ and $\phi$ are called \emph{amplitude} and \emph{phase} of
the coherent pulse, respectively.

In general, the phase of a pulse is completely unknown, or can be
rendered random thanks to a phase randomisation technique. Since the
phase is then uniformly distributed, a pulse state in a given
polarisation is described by the density matrix:
\begin{eqnarray}
\rho_{\mbox{source}} &=&
\frac{1}{2\pi}\int_0^{2\pi}\proj{|\alpha|e^{i\phi}} d\phi\\ &=&
\frac{1}{2\pi}\int_0^{2\pi}
e^{-|\alpha|^2}\sum_{j,j'=0}^{\infty}\frac{|\alpha|^{j+j'}}{\sqrt{j!j'!}}e^{i\phi(j-j')}\ket{j}\bra{j'}d\phi\\
&=&
\sum_{j=0}^{\infty}e^{-|\alpha|^2}\frac{|\alpha|^{2j}}{j!}\proj{j}\label{Fock}
\end{eqnarray}
Therefore, the signals emitted by a coherent source of light becomes a
classical mixture of Fock states due to the lack of a phase
reference. Another example of practical source is a source emitting
thermal states of light. Such states are already mixtures of Fock
states. The above de-phasing argument applies in general for any
signal state. Further studies of source characterisation can be found
in~\cite{walls94a}.

We summarise the assumptions on the quantum setup adopted throughout
this paper:
\begin{itemize}
\item The legitimate parties use a source of photons that sends a
mixture of Fock states $\rho =\sum_{n=0}^\infty p_n\proj{n}$ in the
polarisation state exactly as specified by the user. The numbers of
photons in the pulses emitted by the source are assumed to be
identically and independently distributed. The upper bound $\Mmax$ on
the emitted number of multi-photon signals during the protocol is
known by the legitimate parties to hold except with a negligible
probability $Pr(M>\Mmax)$.
\item The legitimate parties use a photo-detection unit such that for
any given signal, the probability of detection is independent of the
choice of the measurement basis.
\item The signals and Alice's and Bob's polarization bases are chosen
truly at random.
\item Eve cannot intrude Alice's or Bob's apparatus by utilizing the
quantum channel. She is restricted to interaction with the signals as
they pass along the quantum channel.
\end{itemize}

%%%%%%%%%%%%%%%%%%%%%%%%%%%%%%%%%%
%	3.2 The protocol
%%%%%%%%%%%%%%%%%%%%%%%%%%%%%%%%%%

\subsection{The protocol}\label{Protocol}

The quantum key distribution protocol under consideration based on
Bennett and Brassard's BB84~\cite{BB84} is defined. It comprises three
stages: agreement on parameters of the protocol and security
constants, the transmission of quantum signals, and the execution of a
classical protocol together with the validation test.

\begin{description}
\item[Pre-agreement]\hfill
\begin{enumerate}
\item Alice and Bob specify:
\begin{itemize}
\item $m$, the length (in bits) of the private key to be generated.
\item $N$, the number of quantum signals to be sent by Alice. This
integer is the security parameter of the protocol.
\item $\delta$, the maximum threshold value for the error rate for the
validation test.
\item $\rmin$, the minimum threshold value for Bob's detection rate
($1 > \rmin > \Mmax/N$).
\item $p_R$, the proportion of the shared bits that must be publicly
announced for the validation test ($0 < p_R \leq 1/2$).
\item $\tau_{ec}$, $\tau_f$, $\tau_M$, $\tauh$, and $\tau_p$ the
security constants of the protocol. They are small strictly positive
real numbers chosen so that $\delta+\tau_{ec} < 1$, $\delta+\tau_f <
1$, $\rmin > \Mmax/N$, $\tauh < \frac{1-p_R}{2}$, $\tau_p <1$.
\end{itemize}
\end{enumerate}
\item[Quantum Transmission] \hfill
\begin{enumerate}
\setcounter{enumi}{1}
\item Alice and Bob initialise the counter of the signals as $i=0$ and
Bob initialises the set of detected signals as $\D=\{\}$. Then until
the pre-agreed number of signals have been sent ($i=N$), the following
is repeated\label{transmit}
\begin{enumerate}
\item Alice and Bob increment $i$ by one.
\item Alice picks randomly with uniform distribution a basis
$a_i\in\base$ and a bit value $g_i\in\bin$.\label{stepa}
\item Alice makes her source emit a pulse of photons in the state
$\state{g_i, a_i}$ where $\state{0,+}$, $\state{1,+}$,
$\state{0,\times}$ and $\state{1,\times}$ correspond to single photon
states of polarisation angles 0, $\pi/2$, $\pi/4$ and $-\pi/4$,
respectively. We recall that $\{\ket{\Psi(0,+)}, \ket{\Psi(1,+)}\}$
forms an orthonormal basis of $\H_{\textrm{\scriptsize{photon}}}$, the
Hilbert space for single photon polarisation states, and
$\ket{\Psi(0,\times)}=\frac{\ket{\Psi(0,+)}+\ket{\Psi(1,+)}}{\sqrt{2}}$,
$\ket{\Psi(1,\times)}=\frac{\ket{\Psi(0,+)}-\ket{\Psi(1,+)}}{\sqrt{2}}$.
\item Bob measures Alice's pulse in the basis $b_i$ where
$b_i\in\base$ is chosen randomly at each time.   If at least
one photon is detected, the index $i$ is added to the set $\D$ of
detected signals' indexes, and the outcome of the measurement is
recorded as $h_i\in\bin$ (if the detection unit finds photons in both
modes $h_i=0,1$, the value for $h_i$ is chosen randomly in $\bin$ by
Bob). If no photon is detected at all, $h_i$ is assigned the value
$\perp$.
\end{enumerate}
\end{enumerate}
Note that the random choice of basis in step (d) might be provided by
a beamsplitter (or a coupler) followed by two measurement setups, each
measuring the photons in the basis $+$ and $\times$ respectively. It
might also be given by an external random number generator acting on a
polarisation rotator. 

\item[Classical part] \hfill

We denote by $n$ the number of signals detected by Bob, i.e.~$n=|\D|$,
and by $\a=(a_1,\ldots,a_N)\in\base^N$,
$\b=(b_1,\ldots,\b_N)\in\base^N$, $\g=(g_1,\ldots,g_N)\in\bin^N$ and
$\h=(h_1,\ldots,h_N)\in\{0,1,\perp\}^N$ the outcome of the quantum
transmission (Step \ref{transmit}). Restrictions of these vectors onto
some specified subset $X\subset\{1,\dots,N\}$ will be denoted by
$\a(X), \b(X),\g(X),\h(X)$.
\begin{enumerate}
\setcounter{enumi}{2}
\item Bob announces the set of detected signals by $\D$ to
Alice.\label{annD}
\item Bob picks up randomly a subset of signals which will be revealed
for the validation test $R\subset\{1,\ldots,N\}$ , where each position
$i\in\{1,\ldots,N\}$ is put in $R$ with probability
$p_R$. \label{creaR}
\item Bob announces the revealed set  $R$ and the measurement basis
of all signals $\b$ to Alice.\label{announceb}\label{annRb}
\item Bob announces the bit values of the test set $\h(\D\cap R)$ to
Alice.\label{annh}
\item Alice computes the set of corresponding signals $\Omega
=\{i\in\D \st a_i=b_i\}$, the set of corresponding test signals
$T=\Omega\cap R$ and the set of untested corresponding signals
$E=\Omega\cap\overline{R}$. We denote $|E|$ by $\l$.
\item Alice announces the polarisation basis of all of her signals
$\a$, thus announces implicitly $\Omega$ and $E$ as well. The
bitstreams $\g(E)$ and $\h(E)$ are usually called \emph{sifted
keys}.\label{anna}
\item Alice chooses a linear error correcting code~\cite{Sha,Wel}
capable of correcting $\lceil (\delta+\tau_{ec})(1-p_R)|\Omega|\rceil$
errors in $E$. Its parity check matrix, $F$, is a $r\times\l$ binary
matrix, where $r$ is the number of redundant bits required to correct
$\lceil (\delta+\tau_{ec})(1-p_R)|\Omega|\rceil$ errors in $l$ bits
using the linear error correcting code. Alice announces the
\emph{syndrome} $\s=F \g(E)\pmod{2}$ to Bob.
\item Receiving the parity check matrix $F$ and the syndrome $\vec s$,
Bob runs the error correction on his sifted key $\h(E)$ and obtains
$\h'(E)$. If there are less than $\lceil
(\delta+\tau_{ec})(1-p_R)|\Omega|\rceil$ errors in $E$, Bob corrects
successfully all the errors and obtains $\g(E)$, i.e.~$\h'(E)=\g(E)$.
\item Alice picks up randomly with uniform distribution a $m\times\l$
binary matrix $K$ to which we will refer as the {\em privacy
amplification matrix}. Alice announces $K$ publicly.
\item Receiving the privacy amplification matrix $K$, Bob computes
$\k' = K \h'(E) \pmod{2}$.
\end{enumerate}
\item[Validation test] \hfill

Alice runs the validation test. 
\begin{enumerate}
\setcounter{enumi}{12}
\item Alice tests whether the following conditions are all satisfied:
\begin{itemize}
\item Bob's detection rate is greater than $\rmin$, i.e.
\eq{
n>\rmin N.
}
\item The size of $\D$ complies to the following inequalities:
\begin{eqnarray}
\frac{\lh_{min}}{2}&\geq&(\delta+\tau_f)(1-p_R)n ,
\label{argumenthalf}\\
 m+r &\leq&
\lh_{min}\left[1-H_1\left[\frac{2(\delta+\tau_f)\frac{1-p_R}{2}
n}{\lh_{min}}\right]-\tau_p\right]
\label{rateconstraint},
\end{eqnarray}
where 
\eq{
\lh_{min}=\left(\frac{1-p_R}{2}-\tauh\right)(n-\Mmax) }
 is a
probabilistic lower bound on the number of signals on the set $E$
which is due to single photon signals.
\item The number of errors in the tested set $T$ is lower than the
maximally allowed value. More precisely, \eq{ \left|\{i\in T \st
g_i\neq h_i\}\right| < d, } where $d=\delta |\Omega|
p_R$.  
\end{itemize}

The validation test is passed if and only if all the conditions above
are satisfied. The private key is the bitstream obtained by Alice as
follows:

\item Alice computes the private key, defined as:
\begin{itemize}
\item $\k= K \g(E) \pmod{2}$ if the validation test is passed,
\item a $m$-bit string $\k$ chosen randomly with uniform distribution
each time the validation test is not passed.
\end{itemize}
\end{enumerate} 
\end{description}

This protocol defines a key regardless whether the validation test is
passed.  The choice of the security constants used in the protocol is
clarified in the following section.

\paragraph{Note:} The matrix $K$ can be prepared in advance,
and Eve could know its form before the transmission of the quantum
signal. More precisely, Alice and Bob could pre-agree on some set of
matrices  $K$ for various values of $m$, and $l$. It is the
special property \ref{dwh} of $F$ and $K$ which is required here, and
which will be introduced and explained in the section
\ref{privamp}. This property is  satisfied automatically if we
choose  $K$ as random binary matrix, as specified in the
protocol, and the constraint of Eq.~\ref{rateconstraint} is
satisfied. Our security proof can therefore immediately adapted to
other choices of $F$ and $K$ together with their respective
constraints replacing Eq~\ref{rateconstraint} to satisfy the
underlying required property \ref{dwh} of section \ref{privamp}.

%%%%%%%%%%%%%%%%%%%%%%%%%%%%%%%%%%%%%%%%%%%%%%%%%%%%%%%%%%%%%%%%%%%%%%
%	4		 SECURITY OF THE PROTOCOL
%%%%%%%%%%%%%%%%%%%%%%%%%%%%%%%%%%%%%%%%%%%%%%%%%%%%%%%%%%%%%%%%%%%%%%

\section{Security of the protocol}

In this section we present the security statement for the protocol
described in \ref{Protocol}. If follows the structure of Def. 1.  The
proof of the security statement is given in the remainder of the
paper.

\begin{theorem}
The expected conditional Shannon entropy of the key $\rv{\k}$ returned by the
protocol described in Section \ref{Protocol} given Eve's view $\rv{v}$
is lower bounded, for any $N>0$, by
\begin{equation}
H(\rv{\k} | \rv{v}) \geq m -\epsilon_1(N,m)
\end{equation}
where the difference $\epsilon_1(N,m)$ between the bound and the maximal
value $m$ is given by
\begin{eqnarray}
\epsilon_1(N,m) &=& 2 \left( m+\frac{1}{\ln 2} \right) \hd + 2\sqrt{ 2
\left( m+\frac{1}{\ln 2} \right) m \hd} + \nonumber\\
&+&m\Big(e^{-2\tau_M^2 N} + e^{-2\tauh^2 (\rmin N -\Mmax)}+
2^{-\tau_p\left(\frac{1-p_R}{2}-\tauh\right)(\rmin N
-\Mmax)}+\nonumber\\ &+&\sqrt{\gd}\Big).
\end{eqnarray}
where
\begin{eqnarray}
\gd &=& \exp\left[-\frac{1}{2\delta+\tau_f}\tau_f^2 \frac{p_R^2}{4}
\rmin N + 2\left(\frac{\tau_f}{2\delta+\tau_f}\right)^2\right], \\ \hd
&=& 2 \sqrt{\sqrt{\gd}} + \sqrt{\gd}.
\end{eqnarray}

Besides, the conditional probability that Alice and Bob share an
identical private key given that the validation test is passed is
lower bounded for any $N>0$ by: \eq{ \Pr(\neg\share \mbox{ \rm and }
\valid) \leq \epsilon_2(N,m), } where \eq{ \epsilon_2(N,m)
=\min_{\tau_\Omega \in (0,1/2)}\left[
e^{-\frac{1}{2\delta+\tau_{ec}}\tau_{ec}^2p_R^2(\frac{1}{2}-\tau_\Omega)
\rmin
N+2\left(\frac{\tau_{ec}}{2\delta+\tau_{ec}}\right)^2}+e^{-2\tau_\Omega^2
\rmin N}\right] \; .  }
\end{theorem}

The functions $\epsilon_1(N,m)$ and $\epsilon_2(N,m)$ decrease
exponentially with $N$, as required by the definition of security
(Definition \ref{Definition}).

The parameters, the number of emitted signals $N$ out of which the key
of length $m$ is created, are chosen in accordance with the
performance of the set-up used for preparation, transmission and
detection of the quantum signals in view of Equation
\ref{rateconstraint}.  As the number of these transmissions goes to
infinity, we can neglect statistical fluctuations of the signal
properties and describe the ratio between detected signals and sent
signals by a detection rate $p_D=n/N$ and $r_{min} = n/N$.  All
security constants $\tau_{ec}$, $\tau_f$, $\tau_p$, $\tilde{\tau}$ and
$\tau_M$ can be chosen to be arbitrarily small, and the asymptotic key
generation rate out of one bit of the sifted key reads is given as the
length of the sifted key over that of the final key in terms of the
observed error rate $\delta$ as \eq{
\label{asymformula}
\frac{m}{l} = \left( 1-\frac{p_M}{p_D}
\right)\left[1-H_1\left(\frac{2\delta}{1-\frac{p_M}{p_D}}\right)\right]-H_1(\delta)
\; .  } Here we used we used asymptotic equalities for the sifted key
length $l\simeq \frac{1-p_R}{2}n$ and $\lceil
(\delta+\tau_{ec})(1-p_R)|\Omega|\rceil \simeq \lceil \delta
l\rceil$. Furthermore, we made use of the Shannon limit~\cite{Sha}
$r(\lceil \delta l\rceil,l)\simeq l H_1(\delta)$.

\begin{figure}[htb] 
\centerline{\psfig{file=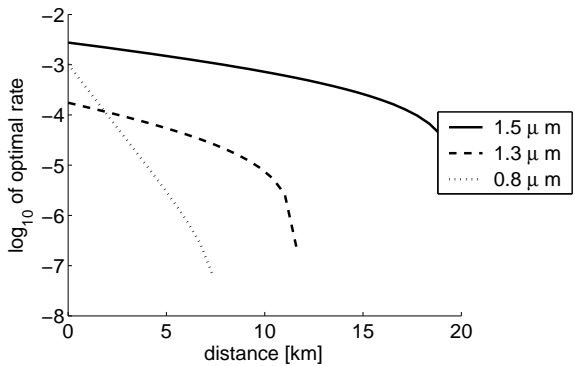,width=3in}}  
\caption{\label{fig1} Asymptotic gain rates using a simulation with the
help of experimental parameters. The parameters are drawn from
Bourennane et al.~\cite{bourennane99a} for $1.5 \mu m$, Marand and
Townsend \cite{marand95a}  for $1.3 \mu m$ and Townsend
\cite{townsend98a} for $0.8 \mu m$.}  
\end{figure}

\begin{figure}[htb] 
\centerline{\psfig{figure=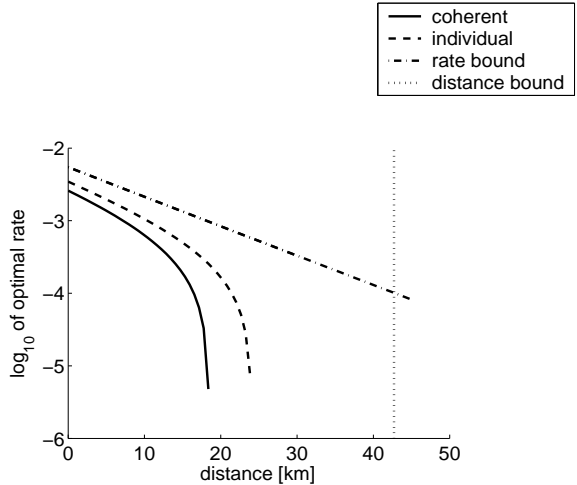,width=3in}}  
\caption{\label{fig3} We use the parameters of
Bourennane et al.~\cite{bourennane99a} for $1.5 \mu m$ to show the
secure gain rate per time slot using our results ('coherent'). For
comparison, the corresponding results for security against individual
attack \cite{nl00a} are given. The rate is bounded due to the
Poissonian photon number distribution of the source and the loss in
the quantum channel ('rate bound') as shown in \cite{nl00a}. The
combination of the source statistics, the loss and detector dark
counts, there is a fundamental bound on the distance over which secure
QKD could be proven with more advances proofs than ours, as shown in
Brassard et al.~\cite{brassard00a} } 
\end{figure}

The overall rate of secure key bits per sent signal $m/N$ can be calculated
directly by multiplying Eq.~(\ref{asymformula}) with the asymptotic formula
\begin{equation}
\frac{l}{N} \simeq \frac{1-p_R}{2} p_D \; .
\end{equation}
The ratio $G$ between key length and received signals $m/n$ can be
obtained by multiplication with $l/n \simeq (1-p_R)/2$. Moreover, in
the limit of arbitrary long keys we can use the limit $p_R \to 0$
since even testing a 'small' fraction of the long key will have
statistical significance sufficient for our purpose. Examples of the
resulting values of $G$ as a function of distance are shown in Figure
\ref{fig1} for various wavelength.
To put our results into context, we relate our results in
Fig.~\ref{fig3} to those obtained for the limited security level of
security against individual attacks. Note that the difference between
the two results is not substantial. More importantly, the difference
might be due to the proof technique used in our result. Our results
should therefore not be interpreted as to claim that coherent attacks
give more information to Eve than individual attacks do.
Furthermore, we lay out the relevant bounds on improved security
proofs. The rate is bounded due to the photon number statistics of the
source, resulting in 
\begin{equation}
G_{bound}=\frac{1}{2} \left(p_D-p_M\right)
\end{equation}
as shown in \cite{nl00a}. We recover this bound by setting $\delta=0$
in our asymptotic bound. 

The distance, over which secure communication is possible, is bounded
by the detector noise. As shown in Brassard et al.~\cite{brassard00a},
the minimal transmission efficiency $F_{WCP}$ in the situation of
Poissonian photon number distribution of the source is given by
\begin{equation}
F_{WCP} \approx 2 \frac{\sqrt{d_B}}{\eta_B}
\end{equation}
where $d_B$ is the dark count probability of the detector per signal
slot and $\eta_B$ is the single photon detection efficiency of the
detector. The corresponding distance (given the parameters of the
experiment) is shown in Fig.~\ref{fig3}.

We have therefore a clear picture of the rates and distances which are
shown to be secure by our proof (the area below the curve 'coherent'
in Fig.~\ref{fig3}), those that are shown to be insecure
\cite{brassard00a,nl00a} (the area outside of the two bound
curves). Note that the area between the 'coherent' line and the two
bounds is the area of the unknown. Future classical protocols taking
on the error correction and privacy amplification tasks from our
protocol in a different way (but leaving the quantum transmission and
measurement untouched) and/or improved security proofs can proclaim
more of this area 'secure'.

%%%%%%%%%%%%%%%%%%%%%%%%%%%%%%%%%%%%%%%%%%%%%%%%%%%%%%%%%%%%%%%%%%%%%%%%%%%%%
%	5		PROOF OF THE MAIN RESULT
%%%%%%%%%%%%%%%%%%%%%%%%%%%%%%%%%%%%%%%%%%%%%%%%%%%%%%%%%%%%%%%%%%%%%%%%%%%%%

\section{Proof of the main result}

The structure of the proof follows. In the first section, an important
feature of the distribution of errors during the quantum transmission
is presented. As an immediate consequence we can proof the
integrity of the protocol, meaning that when the validation test is
passed, Bob shares the private key with Alice with high
probability. The second section deals with the multi-photon signals'
issue. It gives an upper bound on the number of bits a spy can get by
an attack called photon number splitting attack. In the third section,
we explore the method of privacy amplification implemented by binary
matrices and taking into account linear error correction tools.  It turns out
that the privacy of the protocol is equivalent to the ``privacy'' in a
\mp. This equivalence is proved in section 5.4, and the corresponding
mathematical model is provided in section 5.5. Finally, the proof of
privacy of the \mp is given.

There are several points where our proof deviates from that of Mayers
\cite{mayers98a}. Most notably this difference can be seen in 5.3
where the deviation between the proofs shows up quantitatively
. However, changes in the protocol (in our protocol the number of
transmitted signals is fixed which are not necessarily all detected,
and not the number of detected signals, as in \cite{mayers96a}) make
it necessary to check in detail that the basic proof idea of Mayers
carries through.

%%%%%%%%%%%%%%%%%%%%%%%%%%%%%%%%%%%%%%%%%%%%%
%	5.1	On the distribution of errors
%%%%%%%%%%%%%%%%%%%%%%%%%%%%%%%%%%%%%%%%%%%%%

\subsection{On the distribution of errors and the proof of integrity}
We start with a property regarding the distribution of errors which is
based solely on basic probability theory. It allows to make statements
on the key derived from the set $E$ based on the counting of errors in
the set $T$. As an immediate application this property allows us to
proof the integrity of the QKD protocol. Note, that in a practical run
of quantum key distribution, we could omit this estimation, since we
can learn the exact number of errors in $E$ during the later stage of
error correction. However, the kind of estimation presented here
serves a second purpose, which is used later on in our proof. This
purpose is to make a statement about the eavesdropping strategy and
its expected error rate from the observed error rate. Let us explain
this by an example: If Eve implements an intercept/resend attack where
she measures Alice's bit in a randomly chosen signal basis and she
resends a state to Bob corresponding to her measurement result, then
she might be lucky an choose always the correct signal basis. In that
(unlikely) event, she would cause no errors while obtaining full
information on the key. Indirectly, the property below quantifies the
idea that the observed numbers of errors will belong to a typical run
of the protocol. 

\begin{property}\label{Hoef}
Let $\S$ be a set of finite size, $s$. Let $C$ be a randomly chosen
subset of $\S$. The random variable giving the choice of $C$ is
denoted by $\rv{C}$. Let $A$ and $B$ be two subsets of $\S$ chosen
randomly as follows:
\begin{enumerate}
\item Each element in $\S$ is put (exclusively) in $A$ or $B$ or
neither of these sets with respective probabilities $p_A$, $p_B$ and
$1-(p_A+p_B)$. That is, the random variables giving the set to which
the indexes in $\S$ belong to are independently and identically
distributed.
\item Furthermore, the random variables giving the set to which
indexes in $\S$ belong to are independent of the random variable
$\rv{C}$.
\end{enumerate}

We denote by $\rv{A}$, $\rv{B}$ the random variables giving the set
$A$ and $B$, respectively.  Then for any positive real numbers
$\delta$, $\epsilon$ such that $0<\delta<\delta+\epsilon<1$, \eq{ \Pr(
|\rv{A}\cap\rv{C}| < \delta s p_A \mbox{ \rm and } |\rv{B}\cap\rv{C}|
\geq (\delta+\epsilon)s p_B)\leq f(\delta,\epsilon,p_A,p_B,s) } where
\eq{ f(\delta,\epsilon,p_A,p_B,s) =
\exp\left[-\frac{1}{2\delta+\epsilon}\epsilon^2(\min\{p_A,p_B\})^2 s+
2\left(\frac{\epsilon}{2\delta+\epsilon}\right)^2\right].  }
\end{property}

\proof For any subset $C$ of $\S$, given $\rv{C}=C$, each element of
$C$ is either in $\rv{A}$ or in $\rv{B}$ with respective probabilities
$p_A$ and $p_B$.

Now $c=|C|$ is either smaller than
$\lfloor(\delta+\frac{\epsilon}{2})s\rfloor$ or bigger than $\lceil
(\delta+\frac{\epsilon}{2})s \rceil$.
\begin{itemize}

\item If $c \leq \lfloor(\delta+\frac{\epsilon}{2})s\rfloor$, let
$C'=C\cup D$ where $D$ is some subset of $\S\setminus C$ such that
$|C'|=c'=\lfloor(\delta+\frac{\epsilon}{2})s\rfloor$. Then $C \subset
C'$, and \eq{ \Pr (|\rv{B}\cap \rv{C}| \geq (\delta+\epsilon)s p_B |
\rv{C}=C) \leq \Pr (|\rv{B}\cap \rv{C}| \geq (\delta+\epsilon)s p_B |
\rv{C}=C').  }

Furthermore, \eq{ (\delta+\epsilon)s p_B =
\frac{\delta+\epsilon}{\delta+\frac{\epsilon}{2}}p_B
(\delta+\frac{\epsilon}{2})s\geq
(1+\frac{\epsilon}{2\delta+\epsilon})p_B c', } and using the Property
\ref{Binom4} from the Appendix for the set $B$ and the set $C'$,
\begin{eqnarray}
\Pr(|\rv{B}\cap\rv{C}| \geq (\delta+\epsilon) s p_B | \rv{C}=C')
&\leq& \Pr(|\rv{B}\cap\rv{C}| \geq
(1+\frac{\epsilon}{2\delta+\epsilon})p_B c' | \rv{C}=C')\\ &\leq&
\exp\left[-2 \left(\frac{\epsilon p_B}{2\delta+\epsilon}\right)^2 c'
\right]\\ &\leq& f(\delta,\epsilon,p_A,p_B,s),
\end{eqnarray}
since $(\min\{p_A,p_B\})^2\leq p_B^2$ and $c'\geq
\left(\delta+\frac{\epsilon}{2}\right)s-1$. Of course this implies
that \eq{ \Pr(|\rv{A}\cap\rv{C}| < \delta s p_A\mbox{ \rm and }
|\rv{B}\cap\rv{C}|\geq(\delta+\epsilon)s | \rv{C}=C ) \leq
f(\delta,\epsilon,p_A,p_B,s).  }

\item If $c\geq \lceil(\delta+\frac{\epsilon}{2})s\rceil$, then \eq{
\delta s p_A
=\frac{\delta}{\delta+\frac{\epsilon}{2}}p_A(\delta+\frac{\epsilon}{2})s\leq\left(1-\frac{\epsilon}{2\delta+\epsilon}\right)p_A
c } and using the Property \ref{Binom4} for the set $A$ and the set
$C$, \begin{eqnarray} \Pr( |\rv{A}\cap\rv{C}| < \delta s p_A |
\rv{C}=C) &\leq& \Pr( |\rv{A}\cap\rv{C}| <
\left(1-\frac{\epsilon}{2\delta+\epsilon}\right)p_A c | \rv{C}=C)\\
&\leq&
\exp\left[-2c\left(\frac{p_A\epsilon}{2\delta+\epsilon}\right)^2\right]\\
&\leq& f(\delta,\epsilon,p_A,p_B,s),
\end{eqnarray}
since $(\min\{p_A,p_B\})^2 \leq p_A^2 \leq 1$ and $c\geq
\left(\delta+\frac{\epsilon}{2}\right)s >
\left(\delta+\frac{\epsilon}{2}\right)s-1$. Again, this implies that
\eq{ \Pr(|\rv{A}\cap\rv{C}| < \delta s p_A\mbox{ \rm and }
|\rv{B}\cap\rv{C}|\geq(\delta+\epsilon)s | \rv{C}=C ) \leq
f(\delta,\epsilon,p_A,p_B,s).  }
\end{itemize}

We conclude that for any $C$, \eq{ \Pr(|\rv{A}\cap\rv{C}| < \delta s
p_A\mbox{ \rm and } |\rv{B}\cap\rv{C}|\geq(\delta+\epsilon)s |
\rv{C}=C ) \leq f(\delta,\epsilon,p_A,p_B,s).  } Thus
\begin{eqnarray}
\lefteqn{\Pr(|\rv{A}\cap\rv{C}|<\delta s p_A\mbox{ \rm and
}|\rv{B}\cap\rv{C}|\geq(\delta+\epsilon)s p_B)} \nonumber\\ &=&
\sum_{C}\pr{C}(C)\Pr(|\rv{A}\cap\rv{C}|<\delta s p_A\mbox{ \rm and }
|\rv{B}\cap\rv{C}|\geq(\delta+\epsilon)s p_B | \rv{C}=C )\\ &\leq&
f(\delta,\epsilon,p_A,p_B,s),
\end{eqnarray}
which concludes the proof.\fin

An immediate consequence of property \ref{Hoef} is that the error rate
in the sifted key is not significantly higher than the error rate
observed by Alice and Bob during the validation test. This implies the
integrity of the protocol, as defined in Def.~\ref{Definition}, or
more formally:

\begin{property}
The joint probability that Alice and Bob fail to share an identical
key and that the validation test is passed is lower bounded by: \eq{
\Pr(\neg\rv{\share} \mbox{ \rm and } \rv{\valid}) \leq \epsilon_2(N,m)
} where \eq{ \epsilon_2(N,m) = \min_{\tau_\Omega \in (0,1/2)}
\left[e^{-\frac{1}{2\delta+\tau_{ec}}\tau_{ec}^2p_R^2(\frac{1}{2}-\tau_\Omega)\rmin
N+2\left(\frac{
\tau_{ec}}{2\delta+\tau_{ec}}\right)^2}+e^{-2\tau_\Omega^2 \rmin
N}\right].  }
\end{property}
\proof We have seen that Alice and Bob run an error-correcting scheme
capable of correcting $\lceil (\delta+\tau_{ec})(1-p_R)|\Omega|\rceil$
errors in $E$. Thus Bob shares exactly the same key after the error
correction step if there are less than
$(\delta+\tau_{ec})(1-p_R)|\Omega|$ errors in $E$. Given that
$\rv{\Omega}=\Omega$ where $\Omega\subset \{1,\ldots,N\}$, the
probability that the validation test passes while there are more than
$(\delta+\tau_{ec})(1-p_R)|\Omega|$ errors in $E$ is bounded by:
\begin{eqnarray}
\lefteqn{\Pr(\rv{\P}(\rv{T},\delta|\Omega|
p_R)\wedge\neg\rv{\P}(\rv{E},(\delta+\tau_{ec})|\Omega|
(1-p_R)))}\nonumber\\ &=& \Pr( |\rv{T}\cap\rv{C}| < \delta |\Omega|p_R
\mbox{ \rm and } |\rv{E}\cap\rv{C}|\geq
(\delta+\tau_{ec})|\Omega|(1-p_R)) \\ &\leq&
f(\delta,\tau_{ec},p_R,1-p_R,|\Omega|)\nonumber\\ &\leq&
\exp\left[-\frac{1}{2\delta+\tau_{ec}}\tau_{ec}^2p_R^2|\Omega| +
2\left(\frac{\tau_{ec}}{2\delta+\tau_{ec}}\right)^2\right].
\end{eqnarray}
using the above property for $\S=\Omega$ and where $\rv{C}$ is the
random variable giving the set of discrepancies between Alice's bits
$\g(\Omega)$ and Bob's bits $\h(\Omega)$ on $\Omega$. Indeed, $\rv{R}$
is independent of $\rv{C}$, and consequently the random variables
giving the set ($E$ or $T$) to which the indexes in $\Omega$ belong to
are independently and identically distributed ($\Pr( i\in E | i\in
\Omega)=(1-p_R)$ and $\Pr(i\in T | i\in\Omega)=p_R$), and independent
of $\rv{C}$. The above implies that the probability that the error
correction fails to reconcile Alice's and Bob's sifted keys while the
validation test is passed is upper-bounded by an exponentially
decreasing function of $|\Omega|$. Now, each index in $\D$ has
probability $1/2$ to be put in the set $\rv{\Omega}$. Let
$\tau_\Omega$ be a constant obeying $0 < \tau_\Omega <1/2$. Suppose we
are given that $\rv{n}=n$ for some positive integer $n$. Using
Property \ref{Binom4} in the Appendix, the probability that there are
less than $(\frac{1}{2}-\tau_\Omega)n$ is bounded by: \eq{
\Pr(|\rv{\Omega}| \leq (\frac{1}{2}-\tau_\Omega)n \,|\, \rv{n}=n)\leq
e^{-2\tau_\Omega^2 n}.  }

Therefore,
\begin{eqnarray}
\lefteqn{\Pr(\neg \rv{\share} \wedge \rv{\valid})}\nonumber\\ &\leq&
\Pr\left(\rv{\P}(\rv{T},\delta|\rv{\Omega}|
p_R)\wedge\neg\rv{\P}(\rv{E},(\delta+\tau_{ec})|\rv{\Omega}|
(1-p_R))\,\Big|\, \rv{n} > \rmin N \right) \\ &\leq&
\Pr\left(\rv{\P}(\rv{T},\delta|\rv{\Omega}|
p_R)\wedge\neg\rv{\P}(\rv{E},(\delta+\tau_{ec})|\rv{\Omega}|
(1-p_R))\,\Big|\,|\rv{\Omega}|\geq (\frac{1}{2}-\tau_\Omega)\rv{n}\;,
\; \rv{n} > \rmin N\right)+\nonumber\\ &+& \Pr\left(|\rv{\Omega}| \leq
(\frac{1}{2}-\tau_\Omega)\rv{n} \,\Big|\, \rv{n} > \rmin
N\right)\nonumber\\ &\leq&
e^{-\frac{1}{2\delta+\tau_{ec}}\tau_{ec}^2p_R^2(\frac{1}{2}-\tau_\Omega)\rmin
N+2\left(\frac{\tau_{ec}}{2\delta+\tau_{ec}}\right)^2}+e^{-2\tau_\Omega^2
\rmin N},
\end{eqnarray}
since $\rv{n} > \rmin N$ if the validation test is passed. Since this
equations has to hold for all values $\tau_\Omega \in (0,1/2)$, we
have especially
\begin{equation}
\Pr(\neg \rv{\share} \wedge \rv{\valid})\leq \min_{\tau_\Omega \in
(0,1/2)}
\left[e^{-\frac{1}{2\delta+\tau_{ec}}\tau_{ec}^2p_R^2(\frac{1}{2}-\tau_\Omega)\rmin
N+2\left(\frac{
\tau_{ec}}{2\delta+\tau_{ec}}\right)^2}+e^{-2\tau_\Omega^2 \rmin
N}\right]\; .
\end{equation}
This concludes the proof.

%%%%%%%%%%%%%%%%%%%%%%%%%%%%%%%%%%%%%%%%%%%%%%%%
%	5.2	On multiple photon signals
%%%%%%%%%%%%%%%%%%%%%%%%%%%%%%%%%%%%%%%%%%%%%%%%

\subsection{On multiple photon signals}

Let $\A=\{1,\ldots,N\}$ be the set of indexes of all signals Alice
sent. Each signal Alice sends contains zero, one or more photons, with
respective probabilities denoted by $p_{\scriptsize{V}}$,
$p_{\scriptsize{S}}$ and $p_{\scriptsize{M}}$. Alice does not know how
many photons she actually emits in each individual pulse. However, a
potential eavesdropper Eve can learn the actual number of emitted
photons without disturbing the quantum signal, thanks to a quantum non
demolition measurement (we assume no technological limitation for the
enemy). Let's denote by $V$, $S$ and $M$ the set of indexes of signals
containing zero, one and more photons, respectively. Therefore, $V\cup
S\cup M=\A$ and the set $V$, $S$ and $M$ are disjoint. We will denote
by $\Sigma=(V,S,M)$ this partition of $\A$. We will deal with the
worst case scenario in which the partition $\Sigma$ is unknown to
Alice, but perfectly known to Eve.

In the following sections, a lower bound on the number of bits in the
sifted key not arising from multi-photon signals (that is $|E\cap
\overline{M}|$) will be required. Most of practical implementations of
quantum key distribution today use a quantum channel with high loss
rate, due to technological limitations. This loss rate must be taken
into account to establish the required lower bound. For, Eve could
replace secretly the quantum channel by a perfect quantum channel
without loss (again, we assume no technological limitation for
Eve). Eve might then stop signals containing only one photon, as long
as the resulting loss rate of the quantum channel does not exceed
significantly the expected loss rate of the original channel. By doing
so, Eve increases the proportion of bits arising from multi-photon
signals in the sifted key, without being noticed by the legitimate
users. Now if a signal sent by Alice contains several photons, Eve can
split off one photon from the pulse without disturbing the
polarisation of the remaining photons. She stores the stolen photon
until bases are announced and learns deterministically the
corresponding bit by measuring it in the correct basis. This attack is
usually referred to as the \emph{photon number splitting}
attack~\cite{brassard00a, nl00a}. It is in view of this attack (in a
slightly different context) that we will need to estimate the number
of bits in the sifted key that are not arising from multi-photon
signals.

It is possible to give a probabilistic lower bound on the number of
bits in the sifted key that are not arising from multiple photon
signals, provided that an upper-bound on the probability $p_M$ is
given. More precisely,

\begin{property}\label{lh}
Let's denote by $\lh$ the number of bits in $E$ that are not arising
from multi-photon signals, i.e. $\lh=|E\cap\overline{M}|$. We denote
by $\rv{\lh}=|\rv{E}\cap\rv{\overline{M}}|$ the corresponding random
variable.  We recall that we defined the random variable
$\rv{\lh_{min}}$ as: \eq{ \rv{\lh_{min}} =
\left[\frac{1-p_R}{2}-\tauh\right](\rv{n} -\Mmax) } where the security
constants $\tau_M$ and $\tauh$ are strictly positive real number such
that $\Mmax/N < \rmin$ and $\frac{1-p_R}{2}-\tauh>0$. Then the joint
probability that $\rv{n} >\rmin N$ and that $\rv{\lh} <
\rv{\lh_{min}}$ is bounded by: \eq{ \Pr(\rv{\lh} \leq \rv{\lh_{min}}
\,\wedge\, \rv{n} > \rmin N) \leq e^{-2\tauh^2\left(\rmin N -
\Mmax\right)}+e^{-2\tau_M^2 N} }
\end{property}

\proof We consider the worst case scenario in which all losses and
errors are caused by Eve's intervention on the quantum
channel. Obviously, in order to minimise $\lh$, Eve intervene in such
a way that $M\subset\D$.

Suppose we are given that Bob detected $\rv{n}=n$ signals and that
$\rv{M}=M$. Then there are at least $n-|M|$ signals in $\D$ that are
not arising from multi-photon pulses. Now, each of these
non-multiphoton signals in $\D$ has probability $\frac{1-p_R}{2}$ of
being put in the set $\rv{E}$. Therefore, the probability that there
are less than $\left[\frac{1-p_R}{2}-\tauh\right](n-|M|)$ signals in
the sifted key not arising from multi-photon signals is bounded by:
\eq{ \Pr\left(\rv{\lh} \leq \left[\frac{1-p_R}{2}-\tauh\right](n-|M|)
\Big| \rv{n}=n, \rv{M}=M\right) \leq e^{-2\tauh^2 (n-|M|)} } using
Property \ref{Binom4} in the Appendix.

Now, the marginal probability that Alice sent more than
$(p_M+\tau_M)N$ multi-photon signals is bounded using Property
\ref{Binom4} in the Appendix: \eq{ \Pr(|\rv{M}|\geq (p_M+\tau_M)N
)\leq e^{-2\tau_M^2 N} } since each signal Alice sends has probability
$p_M$ of being in $\rv{M}$.

Note that $\left[\frac{1-p_R}{2}-\tauh\right](n-|M|)\geq \lh_{min}$
whenever $|M|\leq (p_M+\tau_M)N$. Therefore, given that $\rv{n}=n$,
the probability that there are less than $\lh_{min}$ signals in the
sifted key that were not emitted with several photons is bounded by:
\begin{eqnarray}
\Pr(\rv{\lh} \leq \lh_{min}|\rv{n}=n) &\leq& \Pr(|\rv{M}|\geq
(p_M+\tau_M)N \,|\, \rv{n}=n)+\nonumber\\ &+&\Pr(\rv{\lh} \leq
\lh_{min} \mbox{ and }|\rv{M}|\leq (p_M+\tau_M)N| \rv{n}=n)\\ &\leq&
\Pr(|\rv{M}|\geq (p_M+\tau_M)N \,|\, \rv{n}=n) + e^{-2\tauh^2
(n-(p_M+\tau_M)N)}.
\end{eqnarray}

Multiplying both side by $\pr{n}(n)$ and summing over $n > \rmin N$, we get:
\begin{eqnarray}
\Pr(\rv{\lh} \leq \lh_{min}\wedge \rv{n}>\rmin N) &\leq&
\Pr(|\rv{M}|\geq (p_M+\tau_M)N \wedge \rv{n}>\rmin N)+\nonumber\\ &+&
\sum_{n>\rmin N} e^{-2\tauh^2 (n-(p_M+\tau_M)N)} \pr{n}(n)\\ &\leq&
\Pr(|\rv{M}|\geq (p_M+\tau_M)N )+ e^{-2\tauh^2 (\rmin
N-(p_M+\tau_M)N)}\\ &\leq& e^{-2\tau_M^2 N} + e^{-2\tauh^2 (\rmin N
-\Mmax)},
\end{eqnarray}
which concludes the proof.\fin

%%%%%%%%%%%%%%%%%%%%%%%%%%%%%%%%%%%%%%%%%%%%%%%%
%	5.3 	On privacy amplification
%%%%%%%%%%%%%%%%%%%%%%%%%%%%%%%%%%%%%%%%%%%%%%%%

\subsection{On privacy amplification}\label{privamp}

In this section, diverse notions used in connection with privacy
amplification are defined. In particular, we define $\dwh$, the
minimal weight of a privacy amplification code, used in conjunction
with an error-correcting code and an imperfect source.  Finally, an
important probabilistic lower bound on this weight is proved. This
bound will be used in the last part of the proof. It is this
minimal weight which will keep track of the multi-photon signals. The
changed estimation of the minimum weight is therefore the most
important change of this proof as respect to Mayers proof
\cite{mayers98a}, although other details need to be adapted.

The privacy amplification is specified by a $m\times\l$ binary matrix
$K$. The linear error correction code is specified by a $r\times\l$
binary parity check matrix $F$. We introduce some notations. Let $G$
be the $(r+m)\times \l$ matrix: \eq{ G=\left(\begin{array}{c}
F\\K\end{array}\right).  }

For any matrix $A$, $A_{(i)}$ denotes its $i$-th row and $A^{(i)}$ its
$i$-th column.

Recall that $\lh=|E\cap\overline{M}|$ is the number of signals in $E$
that are not arising from pulses sent with several photons.

Let $\ha{G}$ be the $(r+m)\times\lh$ matrix obtained from $G$ by
removing the columns $G^{(i)}$, $i\in M\cap E$, corresponding to the
multi-photon signals. Equivalently, $\ha{G}$ is the matrix formed by
the $\lh$ columns of $G$ corresponding to signals in
$E\cap\overline{M}$. Let $\check{G}$ be the $(r+m)\times (\l-\lh)$
matrix formed by the $(\l-\lh)$ columns $G^{(i)}$, $i\in E\cap
M$. Similarly, we define $\ha{F}$, $\ha{K}$ obtained from $F$, $K$ by
removing the $\l-\lh$ columns $F^{(i)}, G^{(i)}$, $i\in E\cap M$
respectively. And $\check{F}$, $\check{K}$ are the matrices formed by
the $\l-\lh$ columns $F^{(i)}, G^{(i)}$, $i\in E\cap M$
respectively. Thus \eq{ \ha{G}=\left(\begin{array}{c} \ha{F}\\
\ha{K}\end{array}\right),\quad\, \check{G}=\left(\begin{array}{c}
\check{F}\\ \check{K}\end{array}\right).  }

Let $\ha{\G}$ be the set of linear combinations of rows of
$\ha{G}$. Let ${\ha{\G}}^*$ be the set of linear combinations of rows
of $\ha{G}$ which contain at least one row of $\ha{K}$, i.e.  \eq{
{\ha{\G}}^* = \left\{\sum_{i=1}^{r+m} z_i \ha{G}_{(i)}\pmod{2} \,\st\,
\vec z\in\bin^{r+m}, z_j = 1\textrm{ for at least one }
j\in\{r+1,\ldots r+m\}\right\}.  } We define $\ha{\C}$ as: \eq{
\ha{\C} = \left\{\vec x\in\bin^{\lh}\,\st\, \ha{G}\vec x =\vec
0\right\} = \left(\ha{\G}\right)^\bot.  } Note that
$\ha{\C}^{\bot}=\ha{\G}$.  We define the \emph{minimum weight} of
$\ha{\G}^{*}$ as the integer: \eq{ \dwh = \min_{\vec x\in \ha{\G}^{*}}
w(\vec x).  }

Equivalently, \eq{ \dwh = \min_{\vec u\in\bin^r, \vec
v\in\bin^m\setminus \{\vec 0\}} w(\vec u^T \ha{F}+\vec v^T \ha{K}).  }
The minimum weight is an important characterisation of the combination
of the error correction code matrix $F$ and the privacy amplification
matrix $K$. It denotes the minimum number of signals contributing to
key bits or parities of sets of key bits after taking into account
publicly known parities from the error correction code and the
knowledge from multi-photon signals. We need a probabilistic bound on
this quantity. Here we will derive it for the case of random coding
where $K$ is a random binary matrix, but we would like to point out
that other suitable choices for $K$ are indeed possible, and might
lead to increased performance of the protocol in terms of the yield of
secure bits. The important property to be fulfilled is property
\ref{dwh}.

We approach the bound on $\dwh$ via the following lemma taken directly
from~\cite{mayers98a}:

\begin{lemma}
Let $k$, $a$ and $b$ be positive integers. Let $A$ be any $a\times k$
binary matrix. Let $B$ be a $b\times k$ binary matrix, picked at
random with uniform distribution. We denote by $\rv{B}$ the
corresponding random variable. Let $d_{A B}$ be the minimum weight of
linear combinations of rows of $A$ and $B$ that contain at least one
row of $B$: \eq{ d_{AB} = \min_{\vec u\in\bin^a, \vec
v\in\bin^b\setminus \{\vec 0\}} w(\vec u^T A+\vec v^T B).  } Then for
any positive real number $x$ such that $x/k <1/2$ and for any positive
real number $\tau$, \eq{ \frac{a+b}{k} \leq 1-H_1(\frac{x}{k})-\tau
\quad \Rightarrow \quad \Pr(d_{A\rv{B}} < x)\leq 2^{-\tau k} } where
$H_1$ is the binary entropy function.
\end{lemma}

{\bf Proof of the lemma } Let $C$ be the $(a+b)\times k$ matrix
defined by: \eq{C=\left(\begin{array}{c} A\\B\end{array}\right).  }

Define the real number $R$ as $R=k H_1^{-1}(1-\frac{a+b}{k}-\tau)$
where $H_1^{-1}$ is the inverse function of the restricted bijective
function $H_1\,:\, [0,\frac{1}{2}]\rightarrow [0,1]$. Assume that
$\frac{a+b}{k} \leq 1-H_1(\frac{x}{k})-\tau$. This implies that $x
\leq R$. Let $\B$ be the sphere in $\bin^k$ centred at the zero string
$\vec 0$ and of radius $R$. For $i\in\{1,\ldots b\}$, let's denote by
${q}_i$ the probability that there exists $\vec z\in\bin^{a+i-1}$ such
that $\rv{B}_{(i)}+\sum_{j=1}^{a+i-1}z_j C_{(j)}$ is in $\B$
(equivalently, ${q}_i$ is the probability that the coset
$\rv{B}_{(i)}+\Vect\left(\{C_{(j)}\}_{j\leq a+i-1}\right)$ intersects
$\B$). Then
\begin{eqnarray}
\Pr(d_{A\rv{B}} < x) &\leq& \Pr(d_{A\rv{B}} < R)\\
&=& q_1+ q_2 (1-q_1) + \cdots + q_b\prod_{i=1}^{b-1}(1-q_i)\\
&\leq& \sum_{i=1}^{b} {q}_i,
\end{eqnarray}
since the probability that $d_{AB} <R$ is the probability that, if one
picks successively at random the rows $\rv{B}_{(1)}$,
$\rv{B}_{(2)}$,\ldots, $\rv{B}_{(b)}$, at some step
$i\in\{1,\ldots,b\}$ the set $B_{(i)}+\Vect\left(\{C_{(j)}\}_{j\leq
a+i-1}\right)$ intersects $\B$.

Now, \eq{ \left(\rv{B}_{(i)} + \Vect\left(\{C_{(j)}\}_{j\leq
a+i-1}\right)\right)\cap \B\neq \emptyset \Leftrightarrow
\rv{B}_{(i)}\in\left\{ \vec x + \Vect\left(\{C_{(j)}\}_{j\leq
a+i-1}\right) \st \vec x\in \B\right\}, } where the size of the last
set is upper bound by $|\B|\times|\Vect\left(\{C_{(j)}\}_{j\leq
a+i-1}\right)|$. Since $\rv{B}_{(i)}$ is chosen randomly out of
$2^{k}$ strings,
\begin{eqnarray}
{q}_i &\leq& \frac{|\B|\times|\Vect\left(\{C_{(j)}\}_{j\leq
	a+i-1}\right)|}{2^{k}}\\ &\leq& 2^{a+i-1-k}|\B|,
\end{eqnarray}
and using the binomial tail inequality (Property \ref{Binom1}): \eq{
|\B| = \sum_{q=0}^{\lfloor R \rfloor}\binom{k}{q} \leq 2^{k
H_1\left(R/k\right)} \quad \textrm{ for } \frac{R}{k}\leq\frac{1}{2},
} we find \eq{ {q}_i\leq 2^{a+i-1-k +
k\left(1-\frac{a+b}{k}-\tau\right)} = 2^{-b-\tau k+i-1}, } thus \eq{
\Pr(d_{A\rv{B}} < R)\leq \sum_{i=1}^{b}{q}_i = 2^{-b-\tau
k}\sum_{i=0}^{b-1}2^i\leq 2^{-\tau k}.  }

Therefore, the expected probability that ${d_{A\rv{B}}}\leq R$ is
smaller than $2^{-\tau k}$. Thus, \eq{ \frac{a+b}{k} \leq
1-H_1(\frac{x}{k})-\tau \quad \Rightarrow \quad \Pr(d_{A\rv{B}} <
x)\leq 2^{-\tau k} } which concludes the proof of the lemma.\fin

This bound allows us to prove the following crucial property:
\begin{property}\label{dwh}
Let $\rv{\dwh}$ be the random variable giving the minimum weight
$\dwh$ defined above. Then, given that $\rv{n}=n$ for some positive
integer $n$ and $\rv{\lh}\geq \lh_{min}$, \eq{
\Pr\Big(\frac{\rv{\dwh}}{2} < (\delta+\tau_f)\frac{1-p_R}{2} n
\,\Big|\,\rv{\lh}\geq\lh_{min},\, \rv{n}=n,\, \rv{\valid}= \mbox{True}
\Big) \leq 2^{-\tau_p\lh_{min}} }
\end{property}

{\bf Proof} Given that $\rv{n}=n$ and $\rv{\lh}=\lh\geq\lh_{min}$,
note that the random variable $\rv{\ha{K}}$ is uniformly distributed
and independent of other variables.  Passing the validation test in
the protocol requires that the constraint \ref{rateconstraint} \eq{
\frac{m+r}{\lh_{min}} \leq
1-H_1\left[\frac{2(\delta+\tau_f)\frac{1-p_R}{2}n}{\lh_{min}}\right]-\tau_p
} is satisfied. Since the validation test is passed, especially Eqn.~
(\ref{argumenthalf}), the argument of $H_1(x)$ satisfies
$x<1/2$. Moreover, we have $\frac{m+r}{\lh}\leq \frac{m+r}{\lh_{min}}$
and $1-H_1(\frac{2(\delta+\tau_f)\frac{1-p_R}{2}n}{\lh})-\tau_p \geq
1-H_1(\frac{2(\delta+\tau_f)\frac{1-p_R}{2}n}{\lh_{min}})-\tau_p$.
Therefore, the number of rows of $\ha{F}$ and $\ha{K}$ verify: \eq{
\frac{m+r}{\lh} \leq
1-H_1\left[\frac{2(\delta+\tau_f)\frac{1-p_R}{2}n}{\lh}\right]-\tau_p\;
.  }

We can therefore apply the above lemma for $A=\ha{F}$,
$\rv{B}=\rv{\ha{K}}$, $k=\lh$ and
$x=2(\delta+\tau_f)\frac{1-p_R}{2}n$. We obtain that: \eq{
\Pr\Big(\rv{\dwh} < 2(\delta+\tau_f)\frac{1-p_R}{2}n \,\Big|\,
\rv{\lh}=\lh\geq\lh_{min},\, \rv{n}=n,\, \rv{\valid}=
\mbox{True}\Big)\leq 2^{-\tau_p \lh} } or, \eq{
\Pr\Big(\frac{\rv{\dwh}}{2} < (\delta+\tau_f)\frac{1-p_R}{2}n
\,\Big|\, \rv{\lh}\geq\lh_{min},\,\rv{n}=n,\, \rv{\valid}=
\mbox{True}\Big)\leq 2^{-\tau_p \lh_{min}} } which concludes the proof
of the property.\fin

%%%%%%%%%%%%%%%%%%%%%%%%%%%%%%%%%%%%%%%%%%%%%%%%%%%%%%
%	5.4	Reduction to the modified protocol
%%%%%%%%%%%%%%%%%%%%%%%%%%%%%%%%%%%%%%%%%%%%%%%%%%%%%%

\subsection{Reduction to a modified situation}

In this section, a modified situation of the original protocol is
defined. This modified situation does not correspond to a key
distribution, but nevertheless, a ``key'' is defined at Alice's
side. Surprisingly, the ``privacy'' in the modified situation implies
the privacy of the original protocol, and this implication is proved.

\subsubsection{Equivalence with the modified protocol}
We first describe the \emph{modified protocol} which is similar to the
original protocol, except that Bob measures the photons in the sifted
set $E$ in the wrong bases (therefore Bob does not share the private
key with Alice). We show that the security of the modified protocol is
equivalent to the security of the original protocol.

In the subsequent discussion, we will consider -- without loss of
generality as far as the security of the protocol is concerned -- that
Bob's choice of measurement bases $\b$ and the set $R$ are provided by
a randomising box at Bob's side: the box generates randomly a choice
for $R$ and for $\b$ at the beginning of the protocol. It then
provides Bob with the generated data as required by the protocol, that
is, it gives $\b$ during step \ref{transmit} and $R$ at the step
\ref{creaR} to Bob. We now define the intermediate protocol as
follows. In the intermediate protocol,
\begin{itemize}
\item Alice behaves exactly as in the original protocol.
\item Bob's randomising box generates $R$ and $\b$ as before, but
gives $\tb$ instead of $\b$ to Bob at step \ref{transmit}, where: \eq{
\tilde{b}_i \stackrel{Def}{=} \left\{\begin{array}{l} b_i \textrm{ if
} i\in R\\ \neg b_i \textrm{ if } i\notin R.\end{array}\right.  } The
box announces $R$ to Bob at step \ref{creaR} as in the original
protocol.
\item Bob behaves exactly as in the original situation, except that,
in step \ref{annRb}, after he learned the choice for $R$, he computes
and announces $\b$ rather than $\tb$.
\end{itemize}

Therefore, in the modified protocol, Bob measures Alice's signals in
the bases $\tb$ and announces $\b$. The underlying idea is that the
original and the modified protocols are identical, except that Bob
measures the signals indexed in $\overline{R}$ in the wrong bases
(without actually knowing $R$).  Consequently, Alice's sifted key and
Bob's sifted key are uncorrelated: Bob does not share the key with
Alice. The private key is only defined in Alice's hand. Therefore,
this situation does not describe a key exchange. It is only an
abstract stepping stone towards the proof of unconditional privacy,
thanks to the following property:

\begin{property}
 Whichever strategy a potential eavesdropper Eve chooses, the random
 variable giving jointly Alice's private key and Eve's view has the
 same probability distribution in both protocol.
\end{property}

\proof In the following, we say that a random variable in the original
protocol and the corresponding random variable in the modified
protocol are \emph{indistinguishable} if and only if their probability
distributions are identical. A quantum system whose state is not a
priori known is characterised by an ensemble description. Given a
system having probability $p_i$ to be in the state $\rho_i$ for
$i=1,2,\ldots,k$, its \emph{ensemble description} is the list
$\{(p_i,\rho_i)\}_i$, that is, the list of its possible states
together with the corresponding probabilities. We say that a quantum
system in the original protocol and the corresponding quantum system
in the modified protocol are \emph{indistinguishable} if and only if
their ensemble descriptions are identical.  Throughout the proof of
this property, we consider an arbitrary but fixed strategy adopted by
Eve. By strategy, we mean the algorithm or the ``program'' followed by
Eve to eavesdrop. Therefore, if Eve is given the same input, she will
act identically. We have to prove that the data Eve accesses and the
private key Alice gets in the original protocol and in the modified
protocol are indistinguishable if Eve follows this given
strategy. Recall that in the original protocol, Eve learns the values
of $\D$, $R$, ${\b}$, ${\h}({\D}\cap {R})$, ${\P}({T},d)$, ${\a}$,
${F}$, ${\s}$ and ${K}$ via the public discussions. Eve may also
attempt to eavesdrop the quantum channel. If a pulse contains several
photons, Eve might keep one photon and store it until bases are
announced, thus obtaining deterministically the corresponding bit. Eve
may also entangle a quantum probe $P$ to Alice's single photon
signals, and measure $P$ after public discussions. She might also stop
some single photon signals, leaving pulses in vacuum state to Bob.
Let $(\rv{A},\rv{B},\rv{C},\ldots,\rv{D})$ be a set of random
variables (and/or quantum systems) in the original protocol. Let
$(\rv{A'},\rv{B'},\rv{C'},\ldots,\rv{D'})$ be the set of corresponding
random variables (and/or quantum systems) in the modified
protocol. Note that one can show that the set
$(\rv{A},\rv{B},\rv{C},\ldots,\rv{D})$ is indistinguishable from the
set $(\rv{A'},\rv{B'},\rv{C'},\ldots,\rv{D'})$ by showing successively
that: $\rv{A}$ and $\rv{A'}$ are indistinguishable. Given $\rv{A}$ and
$\rv{A'}$ take the same value (denoted as $\rv{A}=\rv{A'}$), $\rv{B}$
and $\rv{B'}$ are indistinguishable. Given $\rv{A}=\rv{A'}$ and
$\rv{B}=\rv{B'}$, $\rv{C}$ and $\rv{C'}$ are indistinguishable,
etc. Now:
\begin{itemize}
\item The choice for $\a$, $\g$, $\b$ and $R$ are indistinguishable in
both protocol. Given that the choice for $\a$, $\g$, $\b$ and $R$
takes the same values in both protocol, Alice announces the same $\a$
in step \ref{anna} and Bob announces the same $\b$ and $R$ in step
\ref{annRb}.
\item Given that Alice's choice for $\a$ and $\g$ take the same value
in both protocol, Alice's quantum signals are indistinguishable in
both protocol.
\item Given that Alice's quantum signals are in the same state in both
protocols, Eve acts on them in the same manner: the interaction of the
quantum signals with Eve's apparatus and the probe $P$ remains the
same. Thus the resulting quantum signals (disturbed and/or suppressed
by Eve) received by Bob are indistinguishable in both
protocol. Likewise, the resulting states of Eve's apparatus and probe
$P$ are indistinguishable in both protocol. Naturally, after the above
coupling, the density matrix describing $P$ does not depend on Bob's
choice of bases or outcomes of the measurements.
\item We assumed that given a quantum signal, the probability that Bob
detects at least one photon in this signal is independent of his
choice of basis. Therefore, given that Alice's quantum signals are
identical in both protocol, the set of detected signals in the
modified protocol is indistinguishable from the set $\D$ of detected
signals in the original protocol.  Given that the choice for $\b$ and
$R$ is the same in both protocol, since $\tilde{b}_i={b}_i$ for $i\in
{R}$, the measurement outcome ${\alt{h}}_i$ in the modified protocol
is indistinguishable from the ${h}_i$ in the original protocol, for
$i\in {R}$. Therefore Bob's announcement of ${\th}({R}\cap{\D})$ in
the modified protocol is indistinguishable from its counterpart in the
original protocol.
\item As a result, the sets ${\Omega}$, ${T}$ and ${E}$ computed by
Alice in the modified protocol are indistinguishable from the
corresponding sets computed in the original protocol.
\item The above implies that the outcome of the test ${\P}({T},d)$ is indistinguishable in both protocol.
\item In both protocol, Alice's choices for ${K}$ and ${F}$ are
indistinguishable. Given $\g$, $E$ and $F$ take the same value in both
protocol, Alice announces the same syndrome ${\s}$.
\item The private data Eve wishes to discover is the private key ${\k}
= {K}{\g}({E}) \pmod{2}$ in both situation.
\end{itemize}

Therefore, the public announcements, Eve's apparatus and probe, and
Alice's private key are indistinguishable in both protocol.  Thus the
random variables giving the results Eve gets from measuring her
apparatus and probe are indistinguishable in both situation. This
concludes the proof.\fin

\subsubsection{Further reduction}

The previous section has shown that it is sufficient to prove privacy
of the modified protocol to prove that the original protocol is
secure. It turns out that it is simpler to prove security for the
modified protocol since Bob has no information about the private
key. The privacy of the modified protocol can be proved even in the
following situation where:
\begin{itemize}
\item Alice announces generously $\g(\overline{E})$ after she
announces $\a$ in step \ref{anna}, and
\item Bob announces generously $\h(\D)$ in step \ref{annD}
(i.e. before announcement of the revealed set $R$), instead of
announcing $\h(\D\cap R)$ in step \ref{annh}.
\end{itemize}

Of course, this can only weaken the security of the modified protocol,
and the security of the resulting protocol implies the security of the
original protocol.

Provided the randomising box is not corrupted and the random choice of
$R$ and $\b$ are announced honestly in step \ref{annRb} by the box,
the security of the modified protocol can be proved even if we
furthermore assume that Bob is corrupted by Eve. That is, Bob tells
Eve the output $\tb$ of the randomising box in step \ref{transmit} and
Eve and Bob together make the measurement they want on the quantum
signals sent by Alice. Bob then announces $\D$ and $\h(\D)$ as told by
Eve in step \ref{annD}.  Thus we can regard the couple Eve-Bob as a
single enemy, provided that the randomising box is not corrupted and
that the public announcement of $R$ and $\b$ in step \ref{annRb} is
made directly by the box.

Of course, $\th(T)$ should be close enough to $\g(T)$ so that the
couple Eve-Bob passes the test. The eavesdropping fails if Alice
declares $\neg\P(T,d)$. After the public discussion, Eve may execute
another measurement on the residual state of the photons to refine her
information.

\subsubsection{Reduction related to multiple photon signals}

We now present a reduction related to the multiple photon signals. By
assuming that the enemy has full knowledge about the multiple photon
signals prior to any public announcement, this reduction will allow us
to work with a simpler situation in which the enemy is performing a
conditional measurement on single photon signals only.

Since Eve has no technological limitation, we must assume that Eve-Bob
have perfect detectors. We also consider the worst case scenario in
which Eve replaces the quantum channel by a perfect one.  Therefore,
Eve-Bob are cheating when the set $\D$ containing all signals in which
Bob officially detected at least one photon is not equal to $S\cup
M$. Eve-Bob choose the set $\D$ at their convenience, while ensuring
that the observed transmission rate $n/N$ is not significantly lower
than the expected transmission rate.  Now, if Alice emits a signal of
index $i$ with several photons, Eve-Bob may pick up one photon from
the signal and measure it in basis $\tilde{b}_i$, giving the outcome
$\alt{h}_i$. Then they measure the remaining photons in the pulse in
the other basis $\neg \tilde{b}_i$, yielding a result $h'_i$. The bit
$\alt{h}_i$ allows Eve-Bob to pass the test for the index $i$, if
$i\in T$. After announcement of Alice's basis $a_i$, Eve-Bob knows
whether $a_i=\tilde{b}_i$ or $a_i=\neg \tilde{b}_i$. In either case,
Eve-Bob learn deterministically $g_i$ (since $g_i=\alt{h}_i$ if
$a_i=\tilde{b}_i$ and $g_i=h'_i$ if $a_i=\neg \tilde{b}_i$). That is,
for any signal $i$ emitted with several photons, Eve-Bob can learn
deterministically $g_i$ while passing the test for the index $i$ with
certainty, if $i\in T$. In order to take into account this extra
knowledge gained by Eve-Bob from the multi-photon signals, we consider
a slightly worse scenario. We henceforth assume that:
\begin{itemize}
\item In addition to sending the photon pulses exactly as described
previously, Alice's source tells secretly Eve-Bob the partition
$\Sigma=(V,S,M)$, the number of photons $n_i$ in each pulse $i$ in $M$
(collectively denoted by $\vec n(M)$), Alice's bases $\a(M)$ and
Alice's bits $\g(M)$. These secret announcements are made at the same
time as the source emits the quantum signals and we denote them
collectively by $\M=(\Sigma, \vec n(M), \a(M), \g(M))$.
\end{itemize}

Again, this assumption can only weaken the security of the
protocol. Now given $\M$, Eve-Bob can re-create the signals sent by
Alice on $M$. That is, provided Eve-Bob learn $\M$, we can assume that
Eve-Bob receive only photon pulses that are in $S$, without modifying
the security of the protocol.

\bigskip

To summarise, the security of the original key distribution protocol
is implied by the security of the modified protocol in which Bob is
corrupted by Eve and in which:
\begin{itemize}
\item $\M=(\Sigma, \vec n(M), \a(M), \g(M))$ are given secretly to
Eve-Bob during step~\ref{transmit}.
\item Eve-Bob receive only photon pulses that are in $S$.
\item Eve-Bob must announce publicly $\h(\D)$ in step~\ref{annD}.
\item Bob's randomising box is not corrupted and announces publicly
$R$ and $\b$ honestly in step~\ref{announceb}.
\end{itemize}

%%%%%%%%%%%%%%%%%%%%%%%%%%%%%%%%%%%%%%%%%%%%%%%%
%	5.5	Mathematical model of
%		eavesdropping in the modified situation
%%%%%%%%%%%%%%%%%%%%%%%%%%%%%%%%%%%%%%%%%%%%%%%%

\subsection{Mathematical model of eavesdropping in the modified situation}

We define the view of Eve-Bob as the set of all data Eve-Bob acquired
during the modified protocol. The random variable describing this view
is denoted by $\rv{v}$, and takes value in the set of all possible
view values, $\Z$. Following our model, the view $v$ has the following
form: \eq{ \rv{v} = (\rv{\M}, \rv{\D}, \rv{\th}(\rv{\D}), \rv{R},
\rv{P}, \rv{j}) } where
\begin{itemize}
\item $\rv{\M} = (\rv{\Sigma}, \rv{\vec n(M)}, \rv{\a(M)},
\rv{\g(M)})$ is the random variable giving collectively the secret
announcements of Alice's source ($\rv{\Sigma} =
(\rv{V},\rv{S},\rv{M})$),
\item $\rv{P} = (\rv{\a},\rv{\g}(\overline{\rv{E}}),\rv{F},\rv{K},\rv{\s})$ is the random variable giving collectively Alice's public announcements, and
\item $\rv{j}$ is the random variable giving collectively the rest of
classical data Eve-Bob obtain by performing measurements on the
quantum signals. The structure of $\rv{j}$ depends, of course, on
Eve-Bob's attack.
\end{itemize}

Note that from the beginning Eve-Bob learn $\tb$ from the random
number generating box. Since the privacy results in the modified
situation will not depend on $\tb$, we will consider $\tb$ as a
parameter of the protocol, known by everybody. This is why the
corresponding random variable is omitted from $\rv{v}$.

We now present the formalism to describe the whole situation just
after Eve-Bob learn $\M$ from the source, that is before they
determine $\D$. Just after Eve-Bob get an outcome $\rv{\M}=\M$, the
situation is modeled as follows:

 The system as seen by Eve-Bob is described in a Hilbert space
 $\H_{sys} = \H_C\otimes\H_S$ where $\H_C$ is the Hilbert space
 describing the classical data $\a$, $\g$, $R$, $F$, $K$ processed by
 Alice or the randomising box and $\H_S$ is the Hilbert space
 describing single photon signals in $S$.

 We will denote by $\rv{c}=(\rv{\a},\rv{\g},\rv{R},\rv{F},\rv{K})$ the
 random variable giving collectively $\a$, $\g$, $R$, $F$, $K$. Each
 possible value $c=(\a,\g,R,F,K)$ for $\rv{c}$ is represented by a
 state (i.e. a normalised vector) $\ket{c}\in\H_C$ such that the set
 $\{\ket{c}\}_c$ forms an orthonormal basis of $\H_C$. The Hilbert
 space $\H_S$ is $\H_S = \otimes_{i\in
 S}\H_{\scriptsize\textrm{photon}}$. The single photon polarisation
 Hilbert space $\H_{\scriptsize\textrm{photon}}$ has been defined
 previously.

For any quantum system described in a Hilbert space $\H$, the state of
the system is fully defined by a Hermitian non negative matrix $\rho$
of unit trace called the \emph{density operator}. When the system has
probability $p_i$ to be in the state $\ket{\Psi_i}$ for $i=1,2,\ldots
,k$ (we say the system is in a \emph{statistical mixture} of states),
then the corresponding density operator is $\rho=\sum_{i=1}^k
p_i\proj{\Psi_i}$. The result of a general measurement on a system
described in $\H$ can be seen as an outcome of a random variable
$\rv{q}$ where $q$ is the measured physical quantity. A general
measurement $\rv{q}$ on a system described in a Hilbert space $\H$ is
described by a \emph{positive operator valued measure} (POVM
henceforth) $\{(q, F_q)\}_{q\in {\mathcal{Q}}}$ where ${\mathcal{Q}}$
is the set of all possible outcomes for $\rv{q}$. It is a set of
Hermitian non negative operators $F_q$ on $\H$ such that $\sum_{q\in
{\mathcal{Q}}}F_q=\Id_{\H}$. Then the probability that the measurement
yields a particular value $q$ is given by \eq{ \pr{q}(q) =
\Tr(F_q\rho) } where $\rho$ is the density operator of the system.
For any $q\in {\mathcal{Q}}$, the Hermitian nonnegative operator $F_q$
is called the \emph{positive operator} associated with the outcome
$q$. A more detailed description of the general measurement formalism
can be found in~\cite{peres93a}.

This formalism can be applied to our system $\H_{sys}=\H_C\otimes
\H_S$. However, we need to describe $c$ as classically encoded
variable. This is done by adding the following restrictions to the
above formalism:

\begin{itemize}
\item Any state in $\H_C\otimes\H_S$ should be described as a mixture
of states in the canonical or the \emph{computational} basis of
$\H_C$, i.e. its density matrix must be of the form: \eq{ \rho_{sys} =
\sum_{c} \pr{c}(c)\proj{c}\otimes\proj{\Phi_{c}} } where computational
basis means that no other basis than the canonical one
$\{\ket{\a,\g,R,F,K}\}_{c}$ should be used (i.e. we shall not use
basis containing cat-state vectors such as
$\frac{\ket{c_1}+\ket{c_2}}{\sqrt{2}}$). The probability $\pr{c}(c)$
is the probability of occurrence of $c$.

\item Any positive operator describing a general measurement on
$\H_C\otimes\H_S$ should be of the form: \eq{ \Pi^C\otimes E^Q } where
$\Pi^C$ (acting on $\H_C$) is some projection operator on the
computational basis of $\H_C$ (i.e. on the subspace spanned by some
set of vectors of the canonical basis). In other words, \eq{ \Pi^C =
\sum_{c\in A}\proj{c} } for some set $A$ of values $c$ may take. The
set $A$ corresponds to the set of values $c$ that are compatible with
the outcome associated with the positive operator.

The operator $E^Q$ (acting on $\H_{S}$) is some positive operator in
$\H_S$. This model allows global measurement in which two-way
classical communication between Alice and Eve-Bob occurs. This is
necessary since variables such as $\rv{E}$, and $\rv{\P}(\rv{T},d)$
depend on Bob's announcements.
\end{itemize}

In our model, Eve-Bob execute two measurements on the system. The
first one, allowing to find $\D$, $\h(\D)$ given $\M$ but before
public announcement occurs, the second one, allowing Eve-Bob to refine
their information once $P$ is known.

However, technically, it is more convenient to think that Eve-Bob
execute one single POVM measurement on the whole product space
$\H_C\otimes\H_S$. This POVM should obey certain constraints
reflecting the fact that $\D$ and $\th(\D)$ should be measured before
the public announcements by Alice and the box.

Let's now describe more precisely the density matrix of the system and
the POVM associated with various possible measurements during the
protocol.

Once Eve-Bob have learned the value taken by $\rv{\M}$, the density
matrix of the system as seen by Eve-Bob reads, prior to any further
measurement, \eq{ \rho_{|\Mc}=\sum_{c \in
\CM}\pc{c}{\Mc}(c)\proj{c}\otimes\proj{\Psi(\g(S),\a(S))} } where
\begin{eqnarray}
\CM &\stackrel{Def}{=}& \left\{ c'=(\a', \g', R', F', K') \st \a'(M)=
\a(M),\, \g'(M) =\g(M) \right\},\\
\ket{\Psi(\g(S),\a(S))}&\stackrel{Def}{=}&\otimes_{i\in S}
\ket{\Psi(g_i,a_i)}.
\end{eqnarray}
(in the definition of $\CM$, $\a(M)$ and $\g(M)$ are given by
$\M$). The subscript ``$|\Mc$'' stands for ``given $\Mc$''. The
probability distribution of $\pc{c}{\Mc}$ is normalised for each
possible value for the size of $E$, that is, for each possible value
for the number of columns in the matrices $F$ and $K$ (recall that the
size of the parity check matrix and the privacy amplification matrix
is given by the set $E$). This is to ensure that the sum of
probabilities of all outcomes $c=(\a,\g,R,F,K)$ that are compatible
with $|\rv{E}|=n$ is equal to unity, for any possible value $n$. In
other words, $\sum_{F \mbox{ and } K \mbox{have $n$ columns}}
\pc{c}{\Mc}(c)=1$.

Eve-Bob learn the outcome of $\rv{\M}$ which is part of the view
$\rv{v}$. The remaining part of the view is provided by a single
generalised measurement defined by the POVM \eq{ \left\{ \left(v,\,
E_{v|\Mc} \right)\right\}_{v\in\Z_{\M}} } where $\Z_{\M}$ is the set
of views giving $\M$ for the announcement regarding the multiple
photon signals.  We have seen that for any $v\in \Z_{\M}$, $E_{v|\Mc}$
reads \eq{ E_{v|\Mc} = \Pi_{v|\Mc}^C\otimes E_{v|\Mc}^Q } where
$\Pi^C_{v|\Mc}$ is the projection onto the span of states
$\ket{c}\in\H_C$ for all $c$ compatible with the view $v$.

Now $\a$, $R$, $F$ and $K$ are given explicitly by $v$ (of course, the
number of columns in $F$ and $K$ is $|E|$ where $E$ is given by
$v$). The view $v$ tells as well that $\rv{\g}(M) = \g(M)$ (secret
announcement of Alice's source), $\rv{\g}(\overline{E}) =
\g(\overline{E})$ (announcement of $\rv{\g}(\overline{E})$) and
$\rv{F\g}(\rv{E}) = F\rv{\g}(E) =\s$ (announcement of $\s$, and note
that $F$ and $E$ are given by $v$). Therefore, the set of all values
for $\rv{c}$ compatible with $v$ is
\begin{eqnarray}
\lefteqn{\left\{ (\a,\y,R,F,K) \st \y\in\CS\right\} \quad \textrm{
where}}\nonumber\\ \CS &=& \left\{\x\in\bin^N \st \x(\overline{E}\cup
M) = \g(\overline{E}\cup M) \textrm{ and } F\x(E) =\s\pmod{2}\right\}
\end{eqnarray}
that is,
\eq{
\Pi^C_{v|\Mc} = \sum_{\x\in\CS}\proj{\a,\x,R,F,K}.
}

Suppose now that at the end of the protocol, and after Eve-Bob get the
view $v$, Alice announces the key $\k$. Then the POVM associated to
this situation reads \eq{ E_{(v,\k)|\Mc} = \Pi_{(v,\k)|\Mc}^C\otimes
E_{v|\Mc}^Q } where $E_{v|\Mc}^Q$ remains the same, since the
additional data come from Alice's announcement only, after the
attack. The set of all values for $\rv{c}$ compatible with $(v,\k)$ in
this situation is
\begin{eqnarray}
\lefteqn{\left\{ (\a,\y,R,F,K) \st \y\in\CSK\right\} \quad \textrm{
where}}\nonumber\\ \CSK &=& \Big\{\x\in\bin^N \st \x(\overline{E}\cup
M) = \g(\overline{E}\cup M)\nonumber\\ & &\quad \textrm{ and } F\x(E)
=\s\pmod{2}\quad\textrm{ and } K\x(E) = \k\pmod{2}\Big\}
\end{eqnarray}

Therefore, \eq{ \Pi^C_{(v,\k)|\Mc} =
\sum_{\x\in\CSK}\proj{\a,\x,R,F,K}.  } Of course, Alice will not
announce publicly $\k$ during the protocol. The above POVM has just
been derived so that we can compute $\pr{v \k}(v,\k)$, the probability
that Eve-Bob get the view $v$ and that the key takes the value $\k$.

Finally, we can assume that for any $v$, the positive operators
$E_{v|\Mc}^Q$ are of the rank one, i.e.  \eq{ E_{v|\Mc}^{Q} =
\proj{\phi_v} } where $\ket{\phi_v}$ are some vectors in $\H_S$. The
vectors $\ket{\phi_v}$ are in general neither normalised nor
orthogonal. The reasons for this assumption follows: suppose a
positive operator $E_{v_0|\Mc}^Q$ has a rank greater than one, namely:
\eq{ E_{v_0|\Mc} = \sum_{i\in I}\proj{\eta_i} } where the vectors
$\ket{\eta_i}\in\H_S$ are possibly not normalised (such decomposition
is always possible since $E_{v_0|\Mc}$ is Hermitian positive). $I$ is
a set of size greater than 1. Then the modified POVM \eq{ \{(v,
E_{v|\Mc})\}_{v\neq v_0}\cup
\{((v_0,i),\Pi^C_{v_0|\Mc}\otimes\proj{\eta_i})\}_{i\in I} } gives
more precise information than the original POVM. This justifies our
assumption.

Finally, we examine the constraint on the POVM
$\{\left(v,E_{v|\Mc}\right)\}_{v\in\Z_\M}$ related to the fact that
given $\M$, Eve-Bob must determine $\D$ and $\th(\D\cap \overline{M})$
($\g(M)$ is already known and Eve-Bob do not commit error on $M$)
prior to Alice's public announcements. We have seen that Eve-Bob may
choose the set $\D$ at their convenience. Since signals in $M$ give
perfect information about Alice's bits and signals in $V$ give no
information at all, we assume that Eve-Bob follow the optimal strategy
by choosing $\D$ such that: \eq{ M \subset \D\quad\textrm{ and }\quad
\D\cap V=\emptyset }

Now, since $\M$, $\D$ and $\th(\D\cap\overline{M})$ are parts of the
view $v$, we can define the POVM
\begin{eqnarray}
\lefteqn{\left\{\left((\M,\D,\th(\D\cap\overline{M})),
E_{\D,\th(\D\cap\overline{M})|\Mc}\right)\right\}_{\D \st M\subset \D,
\D\cap V=\emptyset} \quad\textrm{ with }}\nonumber\\ &&
E_{\D,\th(\D\cap\overline{M})|\Mc} = \sum_{v\,\textrm{\scriptsize{
gives }} \, \D,\th(\D\cap\overline{M})} E_{v|\Mc}
\end{eqnarray}
which is the positive operator associated with the outcome
$(\rv{\D},\rv{\th}(\rv{\D}\cap\rv{\overline{M}}))=(\D,\th(\D\cap\overline{M}))$
given that $\Mc$.  When Eve-Bob make a measurement to determine $\D$
and $\th(\D\cap\overline{M})$, the only data they have about $\rv{c}$
are $\a(M)$ and $\g(M)$. Therefore, \eq{ E_{\D,\th(\D\cap\overline{M})
| \Mc} = \Pi^C_{\M}\otimes E_{\D,\th(\D\cap\overline{M})|\Mc}^Q }
where $E_{\D,\th(\D\cap\overline{M})|\Mc}^Q$ is some positive operator
acting on $\H_S$ and \eq{ \Pi^C_{\M} = \sum_{c\in \CM} \proj{c}.  }

To recapitulate, for any positive real number $e>0$, the test
$\rv{\P}(\rv{A},e)$ on a subset $\rv{A}$ of $\D$ is modeled as
follows:
\begin{itemize}
\item Eve-Bob get an outcome $\rv{\M}=\M$ for the multiple photon
signals, thanks to Alice's source.
\item Given $\Mc$ Eve-Bob determine the value taken by $\rv{\D}$ and
$\rv{\th}(\rv{\D}\cap\overline{M})$ thanks to the POVM
\eq{\label{test} \left\{\left((\M,\D,\th(\D\cap\overline{M})),
E_{\D,\th(\D\cap\overline{M})|\Mc} = \Pi^C_{\M}\otimes
E_{\D,\th(\D\cap\overline{M})|\Mc}^Q\right)\right\}_{\D \st M\subset
\D, \D\cap V=\emptyset} }
\item Eve-Bob do not commit any error on $A\cap M$.
\end{itemize}

%%%%%%%%%%%%%%%%%%%%%%%%%%%%%%%%%%%%%%%%%%%%%%%
%	5.6	Bound on the conditional entropy
%		of the key in the modified situation
%%%%%%%%%%%%%%%%%%%%%%%%%%%%%%%%%%%%%%%%%%%%%%%

\subsection{Bound on the conditional entropy of the key in the modified situation}\label{privmodif}

In this section, we derive the bound on the conditional entropy of the
key in the modified situation. Throughout this section, we consider a
given eavesdropping strategy chosen by Eve-Bob that fits the model we
gave previously.

The structure of the proof follows. We define the subset $\P$ of views
in which Eve-Bob succeed to pass the validation test (recall that in
our protocol, the outcome of the validation test is publicly
announced). We define two subsets $\L$ and $\R$ of $\P$. The subset
$\L$ is the set of views for which the associated positive operators
obey a certain constraint. This constraint is related to the fact that
it is very unlikely that Eve-Bob pass the validation test while they
have a substantial knowledge about Alice's sifted key: indeed, if a
quantum signal is in the revealed set $R$, Eve-Bob want to learn the
outcome of the measurement in the basis indicated by the randomising
box. If it is not in $R$, then Eve-Bob want to learn the measurement's
outcome in the conjugate basis (since $\tilde{b}_i=\neg b_i$ if
$i\notin R$). The trouble for Eve-Bob is that they do not know $R$
before they have to announce their bits $\h(\D)$ and this can be
translated in the form of the above constraint. The second subset $\R$
corresponds to the set of views in which probabilistic properties we
have seen previously actually hold. We prove useful identities on $\R$
that are necessary in the subsequent part of the proof. We then prove
that: 1) when the view is in the intersection of $\R$ and $\L$,
Alice's private key is almost uniformly distributed and independent of
Eve-Bob's view, and 2) this intersection covers almost completely the
set $\P$ of views passing the test. Then conclusive calculations lead
to the privacy of the protocol.

The following lemma will be useful in this section.

\begin{lemma}\label{Lemma3}
Let the density matrix of the system be of the form: \eq{ \rho_{sys} =
\sum_{c} \pr{c}(c)\proj{c}\otimes\proj{\Phi_c} } where
$\{\ket{\Phi_c}\}_c$ is an orthonormal set of vectors in $\H_S$, and
let a positive operator acting on $\H_{sys}$ be of the form: \eq{ F =
\left(\sum_{c\in A}\proj{c}\right)\otimes F^Q } where $A$ is some set
of values for $c$. Then for any operators $V$ and $W$ acting on
$\H_S$, \eq{ \Tr\left( F V \rho_{sys} W\right) = \pr{c}(A) \Tr\left(
F^Q V\rho_{sys, A} W\right) } provided $\pr{c}(A)>0$, where
\begin{eqnarray} 
\pr{c}(A) &=& \sum_{c'\in A}\pr{c}(c') \mbox{ and}\\
\rho_{sys, A} &=& \frac{1}{\pr{c}(A)}\sum_{c\in A}\pr{c}(c)\proj{\Phi_c}.
\end{eqnarray}
\end{lemma}

\proof
We have:
\begin{eqnarray}
\Tr(F V\rho_{sys} W) &=& \sum_{c\in
A}\sum_{c'}\pr{c}(c')\underbrace{|\langle
c|c'\rangle|^2}_{\delta_{c,c'}} \Tr(F^Q V\proj{\Phi_{c'}} W)\\ & &
\mbox{ where } \delta_{X, X'}=\left\{\begin{array}{cl} 0 & \mbox{ if }
X\neq X'\\ 1 & \mbox{ if } X=X'\end{array}\right.\nonumber \\
&=&\sum_{c\in A}\pr{c}(c)\Tr(F^Q V\proj{\Phi_c} W)\\ &=&\Tr\big(F^Q
V\sum_{c\in A}\pr{c}(c)\proj{\Phi_c} W\big).
\end{eqnarray}

Now if $\pr{c}(A) = \sum_{c'\in A}\pr{c}(c') > 0$, then \eq{ \Tr(F
V\rho_{sys} W)=\pr{c}(A)\Tr\big( F^Q V
\underbrace{\frac{1}{\pr{c}(A)}\sum_{c\in
A}\pr{c}(c)\proj{\Phi_c}}_{=\rho_{sys,A}} W\big).  }

The factor $\pr{c}(A)$ has been only introduced so that $\rho_{sys,
A}$ is normalised: \eq{ \Tr(\rho_{sys,A}) =
\frac{1}{\pr{c}(A)}\sum_{c\in
A}\pr{c}(c)\underbrace{\Tr(\proj{\Phi_c})}_{=1 \,\forall c}=1.  }

This concludes the proof.\fin

%%%%%%%%%%%%%%%%%%%%%%%%%%%%%%%%%%%%%%%%%%%%%%%%%%%%%%%%
%	5.6.1 Small sphere property
%%%%%%%%%%%%%%%%%%%%%%%%%%%%%%%%%%%%%%%%%%%%%%%%%%%%%%%%

\subsubsection{Small sphere property}\label{ssp}

In this section we define $\L$, the set of views passing the test and
for which the associated positive operators obey a certain
constraint. We then prove that $\L$ covers almost completely $\P$.

\begin{definition}
The set $\P$ is defined as the set of all views of Eve in which the
validation test is passed.
\begin{equation}
\P := \{v \in \Z\;  :\; \valid = true\} \; .
\end{equation}
\end{definition}

\begin{definition} For any view
\eq{ v=(\M, \D, \h(\D), R, P, j)\in \Z } where $\M=(\Sigma, \vec n(M),
\a(M), \g(M))$ and $P=(\a,\g(\overline{E}),F,K,\s)$, define the
\emph{partial view} $z$ as \eq{ z=
(\M,\D,\th(\D\cap\overline{M}),\a,R) \mbox{ part of }v.  }
\end{definition}

The partial view describes the data Eve-Bob have after receiving $\M$
and after measurement of $\D$ and $\th(\D\cap\overline{M})$, followed
by announcements of $(\a, R)$ by Alice and the randomising box. Recall
that Eve-Bob do not make any mistake on $M$ thanks to Alice's source,
and that they need only to get $\th(\D\cap\overline{M})$ using the
POVM (\ref{test}). Given any partial view
$z=(\M,\D,\th(\D\cap\overline{M}),\a,R)$, define $\piz{z}$ as the
orthogonal projection operator onto $\Vect(\{\state{\vec j,\tb} |
d_{E\cap \overline{M}}(\vec j,\th)\geq \db\})$ where
$\db=(\delta+\tau_f)\frac{1-p_R}{2}n $ and where $E$, $M$ and
$\th(\D\cap\overline{M})$ are given by the partial view $z$. We have
restricted to $E\cap \overline{M}$ and $T\cap\overline{M}$ because
Eve-Bob do not commit any error on $M$.  We prove now the following
property (referred to as the small sphere property
in~\cite{mayers98a}).

\begin{property}\label{subsetL}
Let the subset of views $\L\subset\P$ be defined by:
\begin{eqnarray}
\lefteqn{\L\stackrel{Def}{=}\Big\{ v\in \P \st} \nonumber\\
&&\pr{\M}(\M)\Tr\big[E_{v|\Mc}\piz{z}\rho_{|\Mc}\piz{z}\big]\leq
\sqrt{\gd}\pr{v}(v)\Big\},
\end{eqnarray}
where
\eq{
\gd = \exp\left[-\frac{1}{2\delta+\tau_f}\tau_f^2 \frac{p_R^2}{4} \rmin N + 2\left(\frac{\tau_f}{2\delta+\tau_f}\right)^2\right].
}
Then the probability weight of $\L$ is lower bounded by:
\eq{
\pr{v}(\L) \geq \pr{v}(\P)-\sqrt{\gd}.
}
\end{property}

\proof Define $\Z_{\rmin}\subset\Z$ as the subset of views for which
the size of $\D$ satisfies the first condition of the validation test,
i.e. $n > \rmin N$ or $\Z_{\rmin} = \{ v\in\Z \st |\D| > \rmin N
\mbox{ where $\D$ is given by $v$.}\}$. Likewise, define $\W_{\rmin}$
as the subset of partial views $z$ for which the size of $\D$
satisfies the condition $n > \rmin N$, that is $\W_{\rmin}=\{z \st
|\D| >\rmin N\}$. We can assume that $\pr{v}(\Z_{\rmin})$ and
$\pr{z}(\W_{\rmin})$ are strictly positive. Otherwise, since $\P$ is
in $\Z_{\rmin}$, this would imply $\pr{v}(\P)=0$ which implies trivial
security of the protocol. Define the positive operator $\pio{z}$ as
the orthogonal projection operator onto $\Vect(\{\state{\vec j,\tb} |
d_{T\cap \overline{M}}(\vec j,\th)\geq \da\})$ where $\da=\delta
\frac{p_R}{2} n $, and where $T$, $M$ and $\th(\D\cap\overline{M})$
are given by $z$ as before. We also define $\bpio{z}$ as $\bpio{z}=\Id
-\pio{z}$.

We first prove that the set of views $\Q$ defined by:
\begin{eqnarray}
\lefteqn{\Q\stackrel{Def}{=}\Big\{ v\in \Z_{\rmin} \st} \nonumber\\
&&\pr{\M}(\M)\Tr\big[E_{v|\Mc}\bpio{z}\piz{z}\rho_{|\Mc}\piz{z}\bpio{z}\big]\leq
\sqrt{\gd}\pr{v}(v)\Big\}.
\end{eqnarray}
has probability bounded from below by:
\eq{
\pr{v}(\Q) \geq (1-\sqrt{\gd})\pr{v}(\Z_{\rmin})
}
  
Let's assume that we are given that $\rv{\D}=\D$ for some set
$\D$. The starting point is the following: as mentioned already,
Eve-Bob do not know Alice's bases $\a$ nor the choice of $R$ during
the quantum transmission. This means that in a fictional situation
$\F$ in which the single photons sent by Alice are in the state
$\state{\g(S),\tb(S)}$ instead of $\state{\g(S),\a(S)}$ (the
classically stored $\a$ remains however unchanged), Property
\ref{Hoef} holds for the subsets $T$ and $E$ of $\D$. Let $\rv{C}$ be
the random variable giving the set of discrepancies between Alice's
bits $\g(\D)$ and Bob's bits $\h(\D)$ on $\D$. Then in such a
situation, the error set $\rv{C}$ is independent of $\rv{\Omega}$ and
$\rv{R}$.  This implies that $\rv{T}$ and $\rv{E}$ are independent of
$\rv{C}$. Using Property \ref{Hoef} for $\S=\D$, $\rv{A} = \rv{T}$,
$\rv{B}=\rv{E}$ and $\rv{C}$ with $p_A=p_T=p_R/2$, $p_B=p_E =
(1-p_R)/2$ (the factor $1/2$ is the probability that $a_i=\tilde{b}_i$
(for $T$) and $a_i\neq \tilde{b}_i$ (for $E$) respectively), we have
\begin{eqnarray}
\lefteqn{\Pr\big(\rv{\P}(\rv{T},\da)\wedge\neg
\rv{\P}(\rv{E},\db\big)|\F,\,\rv{\D}=\D\big)}\nonumber\\ &\leq&
f(\delta,\tau_f,\frac{p_R}{2},\frac{1-p_R}{2},n).
\end{eqnarray}

Multiplying the above relation by $\pr{\D}(\D)$ and summing for all
$\D$ that satisfy $|\D| >\rmin N$, one gets: \eq{ \Pr\big((\rv{n}
>\rmin N)\wedge\rv{\P}(\rv{T},\da)\wedge\neg
\rv{\P}(\rv{E},\db\big)|\F\big) \leq \gd \pr{v}(\Z_{\rmin})\label{B1}
} remarking that $f(\delta,\tau_f,\frac{p_R}{2},\frac{1-p_R}{2},\rmin
N)=\gd$ and that $\pr{v}(\Z_{\rmin})=\sum_{\D \st |\D| >\rmin
N}\pr{\D}(\D)$.

But the lhs.~above reads:
\begin{eqnarray}
\lefteqn{\Pr\big((\rv{n}>\rmin
N)\wedge\rv{\P}(\rv{T},\da)\wedge\neg\rv{\P}(\rv{E},\db)| \F\big)}
\nonumber\\ &=& \sum_{c}\sum_{z'\in \W_{\rmin}} \pr{c}(c)
\pc{\M}{\rv{c}=c}(\M')\pc{z}{\F,\rv{c}=c,\rv{\M}=\M'}(z')
\Pr(\rv{\P}(\rv{T},\da)\wedge\neg\rv{\P}(\rv{E},\db)| \F, \rv{c}=c,
\rv{z}=z')
\end{eqnarray}
where $\M'$ is given uniquely by the partial view $z'=(\M', \D',
\h'(\D'\cap \overline{M'})$. Note that $\rv{c}$ and $\rv{\M}$ are
independent of the event $\F$.

It is easy to see from (\ref{test}) that, given $\Mc$, the POVM
associated with the partial view $z\in \W_\M$ (where $\W_\M$ is the
set of partial views that are compatible with $\Mc$) is: \eq{ \left\{
z=(\M,\D,\th(\D\cap\overline{M}),\a, R), E_{z|\Mc} =
\Pi^C_{\M,\a,R}\otimes
E^Q_{\D,\th(\D\cap\overline{M})|\Mc}\right\}_{z\in\W_\M} } where \eq{
\Pi^C_{\M,\a,R} = \sum_{ F', K', \g' \st \g'(M)=\g(M)} \proj{\a,\g',
R, F', K'} } is the projection onto states giving $\a$, $\g(M)$ and
$R$ for Alice's choice of bases, Alice's bits on $M$ and the
randomising box's choice for the revealed set, respectively.

Using this POVM, we have:
\begin{eqnarray}
\pc{z}{\F,\rv{c}=c,\rv{\M}=\M'}(z') &=&
\Tr\big[E_{z'|\rv{\M}=\M'}\proj{c}\otimes\proj{\Psi(\g(S'),\tb(S'))}\big]\\
&=& \Tr\big[\Pi^C_{\M',\a',R'}\otimes
E^Q_{\D',\h'(\D'\cap\overline{M'}) |
\rv{\M}=\M'}\proj{c}\otimes\nonumber\\ &\otimes&
\proj{\Psi(\g(S'),\tb(S'))}\big]\\ &=& \delta_{\a,\a'} \delta_{R,R'}
\delta_{\g(M'),\g'(M')}\times\nonumber\\ &\times& \Tr \big[
E^Q_{\D',\h'(\D'\cap\overline{M'}) |
\rv{\M}=\M'}\proj{\Psi(\g(S'),\tb(S'))}\big]
\end{eqnarray}
where $S'$, $M'$ and $\g'(M')$ are given by $\M'$, $\a'$, $R'$, $\D'$
and $\h'(\D'\cap\overline{M'})$ are given by $z'$ and $\a$, $R$ and
$\g$ are given by $c$. We recall that $\M'$ is part of $z'$.

Since Eve-Bob do not commit any error on $\rv{M}$,
\begin{eqnarray}
\lefteqn{\Pr(\rv{\P}(\rv{T},\da)\wedge\neg\rv{\P}(\rv{E},\db)| \F,
\rv{c}=c, \rv{z}=z')}\nonumber\\ &=&
\Pr(\rv{\P}(\rv{T}\cap\overline{\rv{M}},\da)\wedge\neg\rv{\P}(\rv{E}\cap\overline{\rv{M}},\db)|
\F, \rv{c}=c,\rv{z}=z')\\ &=& \Pr(d_{T'\cap \overline{M'}}(\g,\th') <
\da \mbox{ and } d_{E'\cap \overline{M'}}(\g,\th')\geq \db) \\ &=&
\Tr\left(\bpio{z'}\piz{z'}\proj{\Psi(\g(S'),\tb(S'))}\piz{z'}\bpio{z'}\right).
\end{eqnarray}
where the sets $T'$, $E'$ and $M'$ are uniquely given by the partial view $z'$.

Note that
\begin{eqnarray}
\lefteqn{\bpio{z'}\piz{z'}\proj{\Psi(\g(S'),\tb(S'))}\piz{z'}\bpio{z'}}\nonumber\\
&=& \left\{\begin{array}{cl}\proj{\Psi(\tb(S'),\g(S'))} &\textrm{ if }
d_{T'\cap \overline{M'}}(\g,\h') < \da\\ & \textrm{and } d_{E'\cap
\overline{M'}}(\g,\th')\geq \db\\ & \\ 0 & \textrm{
otherwise.}\end{array}\right.
\end{eqnarray}

Therefore, the above term can be integrated in the other trace so that:
\begin{eqnarray}
\lefteqn{\Pr\big((\rv{n}>\rmin
N)\wedge\rv{\P}(\rv{T},\da)\wedge\neg\rv{\P}(\rv{E},\db)| \F)}
\nonumber\\ &=&\sum_{c}\sum_{z'\in\W_{\rmin}}\pr{c\M}(c,\M')
\delta_{\a,\a'} \delta_{R,R'} \delta_{\g(M'),\g'(M')}
\times\nonumber\\ &\times&\Tr\big[E^Q_{\D',\h'(\D'\cap\overline{M'}) |
\rv{\M}=\M'} \bpio{z'} \piz{z'}\proj{\Psi(\g(S'),\tb(S'))} \piz{z'}
\bpio{z'} \big]
\end{eqnarray}
but
\begin{eqnarray}
\pr{c \M}(c,\M') &=& \pr{\M}(\M') \pc{c}{\rv{\M}=\M'}(c) \\ &=&
\pr{\M}(\M') {\rm P}_{\rv{\g}(S')}(\g(S')) {\rm P}_{\rv{\a}
\rv{\g}(\overline{S'}) \rv{ R F K} |\rv{\M}=\M'}(\a,
\g(\overline{S'}), R, F, K) \\ &=& \frac{1}{2^{|S'|}} \pr{\M}(\M'){\rm
P}_{\rv{\a} \rv{\g}(\overline{S'}) \rv{ R F K} |\rv{\M}=\M'}(\a,
\g(\overline{S'}), R, F, K)
\end{eqnarray}
since $\rv{\g}(S')$ is uniformly distributed and independent of
$\rv{M}$, $\rv{\a}$, $\rv{\g}(\overline{S'})$, $\rv{R}$, $\rv{F}$ and
$\rv{K}$. Recall that $\Sigma$ is not chosen by Eve-Bob, but randomly
by the source. Therefore,
\begin{eqnarray}
\lefteqn{\Pr\big((\rv{n}>\rmin
N)\wedge\rv{\P}(\rv{T},\da)\wedge\neg\rv{\P}(\rv{E},\db)| \F\big)}
\nonumber\\ &=& \sum_{z'\in\W_{\rmin}} \sum_{\a, \g(\overline{S'}), R,
F, K} \pr{\M}(\M') {\rm P}_{\rv{\a} \rv{\g}(\overline{S'}) \rv{ R F K}
|\rv{\M}=\M'}(\a, \g(\overline{S'}), R, F, K) \delta_{\a,\a'}
\delta_{R,R'} \delta_{\g(M'),\g'(M')}\times\nonumber\\ &\times&
\Tr\big[E^Q_{\D',\h'(\D'\cap\overline{M'}) | \rv{\M}=\M'} \bpio{z'}
\piz{z'}\sum_{\g(S')}\frac{1}{2^{|S'|}} \proj{\Psi(\g(S'),\tb(S'))}
\piz{z'} \bpio{z'} \big].
\end{eqnarray}

The important point to remark is that \eq{
\sum_{\g(S')}\frac{1}{2^{|S'|}}\proj{\Psi(\g(S'),\tb(S'))} =
\frac{\Id_{S'}}{2^{|S'|}} =
\sum_{\g(S')}\frac{1}{2^{|S'|}}\proj{\Psi(\g(S'),\a(S'))}.  }

Therefore, setting back the sum over $\g(S')$ and writing back the
trace over classical spaces in the original form, we obtain:
\begin{eqnarray}
\lefteqn{\Pr\big((\rv{n}>\rmin
N)\wedge\rv{\P}(\rv{T},\da)\wedge\neg\rv{\P}(\rv{E},\db)| \F\big)}
\nonumber\\ &=& \sum_{c}\sum_{ z'\in\W_{\rmin}}\pr{c \M}(c, \M')
\delta_{\a,\a'} \delta_{R,R'} \delta_{\g(M'),\g'(M')}
\times\nonumber\\ &\times&\Tr\big[E^Q_{\D',\h'(\D'\cap\overline{M'}) |
\rv{\M}=\M'} \bpio{z'} \piz{z'} \proj{\Psi(\g(S'),\a(S'))} \piz{z'}
\bpio{z'} \big]\\ &=& \sum_{z'\in\W_{\rmin}} \sum_{c\in
C_{\M'}}\underbrace{\pr{c\M}(c,\M')}_{=\pr{\M}(\M')\pc{c}{\rv{\M}=\M'}(c)}\Tr\big[E_{z'|\rv{\M}=\M'}\proj{c}\otimes\bpio{z'}\piz{z'}\nonumber\\
& &\quad\proj{\Psi(\a(S'),\g(S'))}\piz{z'}\bpio{z'}\big]\\
&=&\sum_{z'\in\W_{\rmin}}\pr{\M}(\M')\Tr\big[E_{z'|\rv{\M}=\M'}\bpio{z'}\piz{z'}\rho_{|\rv{\M}=\M'}\piz{z'}\bpio{z'}\big],
\textrm{ or,}\\
&=&\sum_{z\in\W_{\rmin}}\pr{\M}(\M)\Tr\big[E_{z|\Mc}\bpio{z}\piz{z}\rho_{|\Mc}\piz{z}\bpio{z}\big],
\end{eqnarray}
where $\M$ is given by $z$.

But $E_{z|\Mc} = \sum_{v \,\mbox{\scriptsize gives}\, z} E_{v|\Mc}$
and we get
\begin{eqnarray}
\lefteqn{\Pr\big((\rv{n}>\rmin
N)\wedge\rv{\P}(\rv{T},\da)\wedge\neg\rv{\P}(\rv{E},\db)|
\F\big)}\nonumber\\ &=&
\sum_{v\in\Z_{\rmin}}\pr{\M}(\M)\Tr\big[E_{v|\Mc}\bpio{z}\piz{z}\rho_{|\Mc}\piz{z}\bpio{z}\big]
\end{eqnarray}
where $\M$ and $z$ are given by $v$, and recalling the Inequality
(\ref{B1}), we get \eq{ \sum_{v\in\Z_{\rmin}}
\pr{\M}(\M)\Tr\big[E_{v|\Mc}\bpio{z}\piz{z}\rho_{|\Mc}\piz{z}\bpio{z}\big]
\leq \gd\pr{v}(\Z_{\rmin}).  }

At this point we use the following lemma:
\begin{lemma}\label{lemma1}
Let $\mu$ be a strictly positive real number. Let $\rv{y}$ be a random
variable taking values in a set $\Y$. Let $\{a_y\}_{y\in\Y}$ be a set
of $|\Y|$ real nonnegative numbers such that $\sum_{y\in\Y}a_y\leq
\mu$. Let $q$ be a strictly positive number. If we define the subset
$\X\subset\Y$ by \eq{ \X = \{y\in\Y \st a_y\leq \mu q \pr{y}(y)\} }

Then $\pr{y}(\X)\geq 1-\frac{1}{q}$.
\end{lemma}

\proof Assume to the contrary that the set $S=\Y\setminus\X = \{y\in\Y
\st a_y >\mu q\pr{y}(y)\}$ has probability $\pr{y}(S)$ greater than
$\frac{1}{q}$. Then \eq{ \sum_{y} a_y\geq \sum_{y\in S}a_y > \mu
q\sum_{y\in S}\pr{y}(y)= \mu q\pr{y}(S)\geq \mu }

Therefore $\sum_y a_y >\mu$ which is a contradiction. This concludes the proof.\fin

Define the set of views $\Q$ as:
\begin{eqnarray}
\lefteqn{\Q\stackrel{Def}{=}\Big\{ v\in \Z_{\rmin} \st} \nonumber\\
&&\pr{\M}(\M)\Tr\big[E_{v|\Mc}\bpio{z}\piz{z}\rho_{|\Mc}\piz{z}\bpio{z}\big]\leq
\sqrt{\gd}\pr{v}(v)\Big\}.
\end{eqnarray}
Then applying the above lemma for $\mu=\gd\pr{v}(\Z_{\rmin})$,
$q=1/\sqrt{\gd}$ and the probability distribution on $\Z_{\rmin}$
given by the conditional distribution $\pr{v}(v)/\pr{v}(\Z_{\rmin})$,
we find that \eq{ \pr{v}(\Q) \geq(1-\sqrt{\gd})\pr{v}(\Z_{\rmin}).  }

Thus, for any view $v\in \Q\cap\P$, we have: \eq{
\pr{\M}(\M)\Tr\big[E_{v|\Mc}\bpio{z}\piz{z}\rho_{|\Mc}\piz{z}\bpio{z}\big]\leq
\sqrt{\gd}\pr{v}(v).  }

However, since $v\in\P$ we also have:
\begin{eqnarray}
\lefteqn{\pr{\M}(\M)\Tr(E_{v|\Mc}
\bpio{z}\piz{z}\rho_{|\Mc}\piz{z}\bpio{z})}\\ &=&\pr{\M \a R F
K}(\M,\a,R,F,K)\pr{\g}(\CS)\Tr\big[E_{v|\Mc}^Q
\bpio{z}\piz{z}\nonumber\\
&&\frac{1}{|\CS|}\sum_{\x\in\CS}\proj{\Psi(\x,\a)}\piz{z}
\bpio{z}\big]
\end{eqnarray}
using Lemma \ref{Lemma3}, and since for any $\x\in \CS$ (note that
$a_i=\tilde{b}_i$ for $i\in T$), \eq{
\bpio{z}\proj{\Psi(\x,\a)}\bpio{z}=\proj{\Psi(\x,\a)} } (that is,
$z=(\M,\D,\h(\D\cap\overline{M}),\a,R)$ verifies
$d_{T\cap\overline{M}}(\h,\x)<\da$ for any $\x\in\CS$). Note that
$\piz{z}$ and $\bpio{z}$ commute. Thus we have:
\begin{eqnarray}
\lefteqn{\forall v\in \Q\cap\P}\nonumber\\
\lefteqn{\pr{\M}(\M)\Tr(E_{v|\Mc}
\bpio{z}\piz{z}\rho_{|\Mc}\piz{z}\bpio{z})}\nonumber\\
&=&\pr{\M}(\M)\Tr(E_{v|\Mc} \piz{z}\rho_{|\Mc}\piz{z})\\ &\leq&
\pr{v}(v)\sqrt{\gd}
\end{eqnarray}
since $\piz{z}$ acts only on $\H_{E\cap\overline{M}}$. This proves
that $\Q\cap\P\subset\L$. Therefore the probability of $\L$ is bounded
from below by:
\begin{eqnarray}
\pr{v}(\L) &\geq& \pr{v}(\Q\cap\P)\\
&\geq& \pr{v}(\P) - \pr{v}(\overline{Q}\cap\Z_{\rmin})\\
&\geq& \pr{v}(\P) - \sqrt{\gd},
\end{eqnarray}
which concludes the proof of the small sphere property.\fin

%%%%%%%%%%%%%%%%%%%%%%%%%%%%%%%%%%%%%%%%%%%%%%%%
%	5.6.2	Identities on \R
%%%%%%%%%%%%%%%%%%%%%%%%%%%%%%%%%%%%%%%%%%%%%%%%

\subsubsection{Identities on $\R$}

Here we define another big subset of $\P$, corresponding to the set of
views in which probabilistic assumptions such as $\lh \geq \lh_{min}$,
$\dwh\geq 2(\delta+\tau_f)\frac{1-p_R}{2}n$ holds. We require as well
that for any $v\in \R$, $\pr{v}(v)>0$. Formally,
\begin{eqnarray}
\R=\{v\in\P \st &v& \mbox{ verifies }\nonumber\\
&& \lh \geq \lh_{min},\nonumber\\
&& \dwh\geq 2(\delta+\tau_f)\frac{1-p_R}{2}n,\nonumber\\
&& \pr{v}(v)>0 \}\label{DefR}
\end{eqnarray}
remembering that $\lh$, $\lh_{min}$ and $\dwh$ are all uniquely
defined by Eve-Bob's view $v$.

In the last section of this proof, a bound on the probability of the
set of views $\overline{\R}\cap\P$ will be needed. We have, using
Properties \ref{lh} and \ref{dwh},
\begin{eqnarray}
\pr{v}(\overline{\R}\cap\P) &\leq& \Pr(\rv{\lh} \leq \lh_{min}\wedge
\rv{n}>\rmin N)+\nonumber\\ &+& \Pr\Big(\frac{\rv{\dwh}}{2} <
(\delta+\tau_f)\frac{1-p_R}{2} n \wedge\rv{\lh}\geq\lh_{min}\wedge
\rv{n}>\rmin N\Big)\\ &\leq& e^{-2\tau_M^2 N} + e^{-2\tauh^2 (\rmin N
-\Mmax}+ 2^{-\tau_p\left(\frac{1-p_R}{2}-\tauh\right)(\rmin N -\Mmax}
\end{eqnarray}

We now prove the following properties on $\R$, i.e. for \eq{
v=(\M,\D,\th(\D),R,P,j)\in\R } where $\M=(\Sigma, \vec n(M), \a(M),
\g(M))$, $\Sigma=(V,S,M)$ and $P=(\a,\g(\overline{E}),F,K,\s)$.  This
implies for instance that $\dwh$ verifies $\dwh\geq
2(\delta+\tau_f)\frac{1-p_R}{2}n$ in this section. It might be useful
to realise that the following properties are exactly equivalent to the
properties proved in the original paper~\cite{mayers98a} in which the
sifted keys $\g(E)$ and $\h(E)$ are replaced by the single-photon
encoded sifted keys $\g(E\cap\overline{M})$ and
$\h(E\cap\overline{M})$.

\begin{property}\label{cardC}
\eq{
\forall v\in\R,\,\forall \k\in\bin^m,\quad |\CS|=2^m|\CSK|
}
\end{property}

\proof
We remark that:
\begin{eqnarray}
\CS =\{\x\in\bin^N \st \x(\overline{E}\cup M)&=&\g(\overline{E}\cup M)
\mbox{ and}\nonumber\\ \ha{F}\x(E\cap\overline{M})&=&\s+\ch{F}\g(E\cap
M)\pmod{2}\}
\end{eqnarray}
($+$ and $-$ are equivalent in arithmetics modulo 2), and
\begin{eqnarray}
\CSK =\{\x\in\bin^N \st \x(\overline{E}\cup M)&=&\g(\overline{E}\cup
M) \mbox{ and}\nonumber\\
\ha{F}\x(E\cap\overline{M})&=&\s+\ch{F}\g(E\cap M)\pmod{2},\nonumber\\
\ha{K}\x(E\cap\overline{M})&=&\k+\ch{K}\g(E\cap M)\pmod{2}\}.
\end{eqnarray}

Now, for $v\in\R$, $\dwh>0$, that is, rows of $\ha{K}$ are linearly
independent and each row of $\ha{K}$ is linearly independent of rows
of $\ha{F}$. Therefore $\ha{K}\x(E\cap\overline{M}) =
\k+\ch{K}\g(E\cap M) \pmod{2}$ introduces $m$ additional linearly
independent constraints in $\CS$. Thus $|\CS|=2^m|\CSK|$.\fin

\begin{property}\label{joint}
For any $\k\in\bin^m$ and $v\in\R$, the mutual probability of the
outcome $(v,\k)$ reads: \eq{ \pr{v \k}(v,\k) =
\frac{1}{2^m}\Pmpr\bra{\phiv}\rhosk\ket{\phiv} } where
\begin{itemize}
\item $\Pmpr=\sum_{\x\in\CS}\pr{\M}(\M)\pc{\a \g R F
K}{\Mc}(\a,\x,R,F,K)$ is the probability that Alice announces
$P=(\a,\g(\overline{E}),F,K,\s)$, the box announces $R$ and Eve-Bob
get $\M$ thanks to the photon number splitting attack.
\item $\state{\g(A),\a(A)}=\otimes_{i\in A}\state{g_i,a_i}\in\H_A$ for
any set $A\subset S$, where $\H_A$ stands for the Hilbert space
describing the photons in $A$.
\item $\rhosk =
\frac{1}{|\CSK|}\sum_{\x\in\CSK}\proj{\Psi(\x(E\cap\overline{M}),\a(E\cap\overline{M}))}$
\item $\ket{\phiv} = \langle\Psi(\g(\overline{E}\cup
M),\a(\overline{E}\cup M))\ket{\phi_v}\in\H_{E\cap\overline{M}}$.
\end{itemize}
\end{property}

Note that in the above notation, $\M$, $P$, $R$, $\CS$ and $\CSK$ are
all given by $v$.

\begin{property}\label{marginal}
For any view $v\in\R$ and for any operators $V$ and $W$ acting on the
restricted space $\H_{E\cap\overline{M}}\subset \H_S$, \eq{
\pr{\M}(\M)\Tr(E_{v|\Mc} V\rho_{|\Mc} W) = \Pmpr\bra{\phiv}V\rhos
W\ket{\phiv} } where
\begin{itemize}
\item $\rhos =
\frac{1}{|\CS|}\sum_{\x\in\CS}\proj{\Psi(\x(E\cap\overline{M}),\a(E\cap\overline{M}))}$
\item and other elements defined as previously.
\end{itemize}
\end{property}

\proof Using Lemma~\ref{Lemma3} for $\rho_{|\Mc}$, $E_{(v,\k)|\Mc}$
and $E_{v|\Mc}$, we get (recall that $\M$ is given by $v$)
\begin{eqnarray}
\pr{v\k}(v,\k) &=&\pr{\M}(\M)\pc{v \k}{\Mc}(v,\k)\\ &=&
\pr{\M}(\M)\Tr(E_{(v,\k)|\Mc}\rho_{|\Mc})\\ &=& \pr{\M}(\M)\pc{\a R F
K}{\Mc}(\a,R,F,K)\pr{\g}(\CSK)\Tr(E_{v|\Mc}^Q\rho_{\s,\k,\g(\overline{E}\cup
M)})
\end{eqnarray}
where \eq{ \rho_{\s,\k,\g(\overline{E}\cup M)} =
\frac{1}{\pr{\g}(\CSK)}\sum_{\x\in\CSK}\pr{\g}(\x)\proj{\Psi(\x,\a)} }
and \eq{ \pr{\M}(\M)\Tr(E_{v|\Mc} V\rho_{|\Mc} W) = \pr{\M}(\M)\pc{\a
R F K}{\Mc}(\a,R,F,K)\pr{\g}(\CS)\Tr(E_{v|\Mc}^Q V
\rho_{\s,\g(\overline{E}\cup M)} W) } where \eq{
\rho_{\s,\g(\overline{E}\cup M)} =
\frac{1}{\pr{\g}(\CS)}\sum_{\x\in\CS}\pr{\g}(\x)\proj{\Psi(\x,\a)}.  }

Now $V$ and $W$ act only on $\H_{E\cap\overline{M}}$ and for any
$\x\in\CS$ or $\CSK$, $\x(\overline{E}\cup M)=\g(\overline{E}\cup
M)$. Thus \eq{ \bra{\Psi(\x,\a)}X\ket{\phi_v} =
\bra{\Psi(\x(E\cap\overline{M}),\a(E\cap\overline{M}))}X\ket{\phiv} }
where $X$ is $V$ or $W$. Noting that $\pr{\g}$ is uniform, for any
$\x$, we have $\pr{\g}(\x)/\pr{\g}(\CS)=1/|\CS|$ and
$\pr{\g}(\x)/\pr{\g}(\CSK)=1/|\CSK|$.  Finally we use the identities
$\pr{\g}(\CSK)=\frac{1}{2^m}\pr{\g}(\CS)$ and $\pr{\M}(\M)\pc{\a R F
K}{\Mc}(\a, R, F, K)\pr{\g}(\CS)=\Pmpr$. This concludes the proof.\fin

It follows that the marginal probability of $v\in\R$ reads:
\eq{
\pr{v}(v)=\pr{\M}(\M)\Tr(E_{v|\Mc}\rho_{|\Mc})=\Pmpr\bra{\phiv}\rhos\ket{\phiv}.
}

Finally, for any ket $\ket{\khi}\in\H_{E\cap\overline{M}}$, for any
$\k\in\bin^m$, we denote by $r_{v,\k}(\ket{\khi})$ the ratio: \eq{
r_{v,\k}(\ket{\khi}) =
\frac{\bra{\khi}\rhosk\ket{\khi}}{\bra{\khi}\rhos\ket{\khi}} }
whenever $\bra{\khi}\rhos\ket{\khi}>0$ and $r_{v,\k}(\ket{\khi})=1$
otherwise.

It is easy to see that, for any view $v\in\R$, any key $\k$ and any
ket $\ket{\khi}\in\H_{E\cap \overline{M}}$ such that
$\bra{\khi}\rhos\ket{\khi}>0$
\begin{eqnarray}
\sum_{\k\in\bin^m}r_{v,\k}(\ket{\khi}) &=&
\frac{\bra{\khi}\sum_{\k\in\bin^m}\rhosk\ket{\khi}}{\bra{\khi}\rhos\ket{\khi}}\\
&=& 2^m\label{RVK}
\end{eqnarray}
where we have used the identity $\sum_{k\in\bin^m} \rhosk = 2^m\rhos$
which follows directly from Property \ref{cardC}. The identity
$\sum_{\k\in\bin^m}r_{v,\k}(\ket{\xi})=2^m$ holds for
$\bra{\khi}\rhos\ket{\khi}=0$ as well.

%%%%%%%%%%%%%%%%%%%%%%%%%%%%%%%%%%%%%%%%%%%%%%%%
%	5.6.3	Quasi-independence of 
%		the key and the view on
%		\R\cap\L
%%%%%%%%%%%%%%%%%%%%%%%%%%%%%%%%%%%%%%%%%%%%%%%%

\subsubsection{Quasi-independence of the key and the view on $\R\cap\L$}

We are going to prove in this section that the probability of the
joint event in which Eve-Bob get the view $v$ and Alice gets the key
$\k$ reads, provided $v\in\R\cap\L$, \eq{ \pr{v\k}(v,\k) =
\pi_v+\eta_{v,\k} } where $\pi_v$ is independent of $\k$ and an upper
bound is found on $|\eta_{v,\k}|$.

For any view $v\in\R$ and any key value $\k\in\bin^m$, we have seen
that (Property \ref{joint}), \eq{ \pr{v\k}(v,\k) =
\frac{1}{2^m}\Pmpr\bra{\phiv}\rhosk\ket{\phiv}.  }

Let $\piw{z}$ be the orthogonal projection onto the subspace
$\H_w=\Vect\big\{\state{\vec j,\tb} \,\big| \,
d_{E\cap\overline{M}}(\vec j,\h)\geq \frac{\dwh}{2}\big\}\subset
\H_S$. The minimum weight $\dwh$ has been defined in
Section~\ref{privamp}. As before, the partial view $z$ is specified by
the view $v$. Let $\bpiw{z}=\Id-\piw{z}$. Then $\piw{z}$ and
$\bpiw{z}$ act non trivially only on $\H_{E\cap\overline{M}}$, and
\eq{ \bra{\phiv}\rhosk\ket{\phiv} =
\bra{\phiv}(\piw{z}+\bpiw{z})\rhosk (\piw{z}+\bpiw{z})\ket{\phiv}.  }
Therefore,
\begin{eqnarray}
\pr{v\k}(v,\k) &=& \frac{1}{2^m}\Pmpr\Big[
\bra{\phiv}\bpiw{z}\rhosk\bpiw{z}\ket{\phiv}+\nonumber\\ &+&
\bra{\phiv}\piw{z}\rhosk\ket{\phiv}+\bra{\phiv}\rhosk\piw{z}\ket{\phiv}-\nonumber\\
&-&\bra{\phiv}\piw{z}\rhosk\piw{z}\ket{\phiv}\Big]\label{Dev}.
\end{eqnarray}

We show that the first term in the rhs.~in Equation (\ref{Dev})
corresponds to the term independent of $\k$ and we derive a bound on
the modulus of the remaining terms in the following parts.

\paragraph{The term independent of the key}
\begin{property}\label{kindep}
For any view $v$ in $\R\cap\L$, the first term in the rhs.~of
(\ref{Dev}) is independent of $\k$. This term will be denoted by
$\pi_v$ subsequently, for any $v\in\R\cap\L$. That is,
\eq{\label{epiv} \pi_v\stackrel{Def}{=} \frac{1}{2^m}
\Pmpr\bra{\phiv}\bpiw{z}\rhosk\bpiw{z}\ket{\phiv}.  }
\end{property}

\proof
We need the following identity:

\begin{lemma}
\begin{eqnarray}
\lefteqn{\forall \vec \alpha, \vec \beta\in\bin^{\lh},}\\
&&\bra{\Psi(\vec \alpha,\tb(E\cap
\overline{M}))}\rhosk\state{\vec\beta,\tb(E\cap \overline{M})} =
\frac{1}{2^{\lh}}\times\left\{\begin{array}{l} 0 \textrm{ if }
(\vec\alpha+\vec\beta)\notin \Gh \\
(-1)^{(\vec\alpha+\vec\beta)\cdot\vec\theta} \textrm{ if
}(\vec\alpha+\vec\beta)\in \Gh.\end{array}\right.
\end{eqnarray}
where $\vec\theta$ is a vector in $\bin^{\lh}$ such that
$\ha{G}\vec\theta=\left(\begin{array}{c}\s\\
\k\end{array}\right)+\ch{G}\g(E\cap M)$ ($\vec\theta$ exists since
$|\CSK|>0$ for $v\in\R$). We recall that $\Gh$ has been defined in
Section \ref{privamp}.
\end{lemma}

{\bf Proof of the lemma } First we need some definitions. For
$y\in\bin$ and for $a\in\{+,\times\}$, define the unitary operator
$U^a_y$ acting on a single photon Hilbert space: \eq{ \forall
x\in\bin,\quad U^a_y\ket{\Psi(x,a)} = \ket{\Psi(x+y, a)} } It is easy
to verify that on the opposite basis $U^a_y$ acts as:
\begin{equation}
U^a_y\ket{\Psi(x,\neg a)} = (-1)^{x y}\ket{\Psi(x,\neg a)}. \label{2}
\end{equation}

Likewise, for $\vec y\in\bin^{\lh}$ define the unitary operator $\U_y$
acting on $H_{E\cap\overline{M}}$ as: \eq{ \forall \x\in\bin^{\lh},
\quad \U_{\vec y}\ket{\Psi(\x,\a(E\cap\overline{M}))} =
\ket{\Psi(\x+\vec y,\a(E\cap\overline{M}))}.  }

It is easy to see that $\U_{\vec y}$ is involutive, that is $U_{\vec
y}^{\a(E\cap\overline{M})\,-1} = \U_{\vec y}$. Since $\tilde{b}_i=\neg
a_i$ for $i\in E\cap\overline{M}$, we have, using equation (\ref{2}),
\eq{ \forall \x,\quad \U_{\vec y}
\ket{\Psi(\x,\tb(E\cap\overline{M}))} = (-1)^{\x\cdot\vec
y}\ket{\Psi(\x,\tb(E\cap\overline{M}))}.\label{26} }

Returning to our proof, we express $\rhosk$ (defined in Property
\ref{marginal}), recalling that
$\ha{G}=\left(\begin{array}{c}\ha{F}\\\ha{K}\end{array}\right)$. Furthermore,
we use the fact that for any $\vec y\in\bin^{\lh}$, \eq{ \ha{G}\vec y
= \left(\begin{array}{c}\s\\\k\end{array}\right)+\ch{G}\g(E\cap M)
\Leftrightarrow \y\in\vec\theta+\Ch } where $\theta$ is a vector in
$\bin^{\lh}$ such that
$\ha{G}\vec\theta=\left(\begin{array}{c}\s\\\k\end{array}\right)+\ch{G}\g(E\cap
M)$ (such $\vec\theta$ exists since $\CSK\neq\emptyset)$. This gives,
recalling that $\Ch = \left(\Gh\right)^\bot$,
\begin{eqnarray}
\rhosk &=& \frac{1}{|\CSK|}\sum_{\begin{subarray}{l}\x\in\bin^{N} |
\\\x(\overline{E}\cup M)=\g(\overline{E}\cup M)\\
\ha{G}\x(E\cap\overline{M})=\left(\begin{subarray}{c}\s\\\k\end{subarray}\right)+\\+\ch{G}\g(E\cap
M)\end{subarray}}\proj{\Psi(\x(E\cap\overline{M}),\a(E\cap\overline{M}))}\\
&=& \frac{1}{|\CSK|}\sum_{\vec
y\in\vec\theta+\Ch}\proj{\Psi(\y,\a(E\cap\overline{M}))}\\ &=&
\frac{1}{|\CSK|}\sum_{\vec
y\in\Ch}\proj{\Psi(\y+\vec\theta,\a(E\cap\overline{M}))},
\end{eqnarray}
and, using Equation (\ref{26}), for all $\vec\alpha,\vec\beta\in\bin^{\lh}$,
\begin{eqnarray}
\lefteqn{\bra{\Psi(\vec\alpha,\tb(E\cap\overline{M}))}\rhosk\ket{\Psi(\vec\beta,\tb(E\cap\overline{M}))}}\nonumber\\
	&=&\bra{\Psi(\vec\alpha,\tb(E\cap\overline{M}))}\frac{1}{|\CSK|}\sum_{\y\in\Ch}\U_{\vec\theta}\ket{\Psi(\y,\a(E\cap\overline{M}))}\times\nonumber\\
	&&
	\times\bra{\Psi(\y,\a(E\cap\overline{M}))}\U_{\vec\theta}\ket{\Psi(\vec\beta,\tb(E\cap\overline{M}))}
	\\ &=&
	(-1)^{(\vec\alpha+\vec\beta)\cdot\vec\theta}\bra{\Psi(\vec\alpha,\tb(E\cap\overline{M}))}{\rho}_{0}\ket{\Psi(\vec\beta,\tb(E\cap\overline{M}))},
\end{eqnarray}
where \eq{ \rho_0 = \frac{1}{|\CSK|}\sum_{\vec y\in
\Ch}\proj{\Psi(\vec y,\a(E\cap\overline{M}))} }

Let $q=\dim\Ch$, and $\{\vec\theta_1,\ldots\vec\theta_q\}$ be a basis
of $\Ch$. Let $\Ch^{(j)}$ be the span of
$\{\vec\theta_1,\ldots\vec\theta_j\}$ for $j\in\{1,\ldots q\}$. For
$j\in\{1,\ldots q\}$, define $\rho^{(j)}$ as: \eq{ \rho^{(j)} =
\frac{1}{2^j}\sum_{\x\in\Ch^{(j)}}\proj{\Psi(\x,\a(E\cap\overline{M}))}.
}

We show by induction on $j\in\{0,\ldots q\}$ that \eq{ \forall
\vec\alpha,\vec\beta\in\bin^{\lh},\quad\bra{\Psi(\vec\alpha,\tb(E\cap\overline{M}))}\rho^{(j)}\ket{\Psi(\vec\beta,\tb(E\cap\overline{M}))}=\left\{\begin{array}{ll}1/2^{\lh}
&\textrm{ if }\vec\alpha+\vec\beta\in\Ch^{(j)\bot} \\ 0 & \textrm{ if
}\vec\alpha+\vec\beta\notin\Ch^{(j)\bot}.\end{array}.\right.\label{Ind}
}

For $j=0$, we have $\Ch^{(0)}=\{\vec 0\}$ and $\Ch^{(0)\bot}
=\bin^{\lh}$ and $\rho^{(0)}=\proj{\Psi(\vec
0,\a(E\cap\overline{M}))}$. Thus \eq{
\forall\vec\alpha,\vec\beta,\quad\bra{\Psi(\vec\alpha,\tb(E\cap\overline{M}))}\rho^{(0)}\ket{\Psi(\vec\beta,\tb(E\cap\overline{M}))}=\frac{1}{2^{\lh}},
} and (\ref{Ind}) holds (Recall $a_i=\neg \tilde{b}_i$ on
$E\cap\overline{M}$).

Suppose (\ref{Ind}) holds for some $j\in\{0,\ldots q-1\}$. Since
$\Ch^{(j+1)}=\Ch^{(j)}\cup(\vec\theta_{j+1}+\Ch^{(j)})$, we have
\begin{eqnarray}
\rho^{(j+1)} &=&
	\frac{1}{2}\Big(\frac{1}{2^j}\sum_{\x\in\Ch^{(j)}}\proj{\Psi(\x,\a(E\cap\overline{M}))}+\nonumber\\
	&&
	+\frac{1}{2^j}\sum_{\x\in\vec\theta_{j+1}+\Ch^{(j)}}\proj{\Psi(\x,\a(E\cap\overline{M}))}\Big)\\
	&=& \frac{1}{2}\big(\rho^{(j)}+\U_{\vec\theta_{j+1}}\rho^{(j)}
	\U_{\vec\theta_{j+1}}\big).
\end{eqnarray}

Thus,
\begin{eqnarray}
\lefteqn{\forall \vec\alpha,\vec\beta,\quad
\bra{\Psi(\vec\alpha,\tb(E\cap\overline{M}))}\rho^{(j+1)}\ket{\Psi(\vec\beta,\tb(E\cap\overline{M}))}}\nonumber\\
&=&\frac{1}{2}\bra{\Psi(\vec\alpha,\tb(E\cap\overline{M}))}\rho^{(j)}\ket{\Psi(\vec\beta,\tb(E\cap\overline{M}))}\underbrace{\big(1+(-1)^{(\vec\alpha+\vec\beta)\cdot\vec\theta_{j+1}}\big)}_{=\left\{\begin{array}{ll}
2 &\textrm{ if }\vec\alpha+\vec\beta\in\vec\theta^{\bot}_{j+1}\\ 0
&\textrm{ if
}\vec\alpha+\vec\beta\notin\vec\theta^{\bot}_{j+1}.\end{array}\right.}.
\end{eqnarray}

And since (\ref{Ind}) holds for $j$, we get \eq{
\bra{\Psi(\vec\alpha,\tb(E\cap\overline{M}))}\rho^{(j+1)}\ket{\Psi(\vec\beta,\tb(E\cap\overline{M}))}=\left\{\begin{array}{ll}1/2^{\lh}
& \textrm{ if }\vec\alpha+\vec\beta\in\Ch^{(j+1)\bot},\\ 0 & \textrm{
if }\vec\alpha+\vec\beta\notin\Ch^{(j+1)\bot}.\end{array}\right.  }
which concludes our induction. Noting that $\Ch^{(q)}=\Ch$,
$|\Ch|=|\CSK|$, $\Ch^{\bot}=\Gh$, and $\rho^{(q)}=\rho_0$, for any
$\vec\alpha$, $\vec\beta \in\bin^{\lh}$, \eq{
\bra{\Psi(\vec\alpha,\tb(E\cap\overline{M}))}\rhosk\ket{\Psi(\vec\beta,\tb(E\cap\overline{M}))}
=\frac{1}{2^{\lh}}\times\left\{\begin{array}{l} 0 \textrm{ if }
(\vec\alpha+\vec\beta)\notin \Gh \\
(-1)^{(\vec\alpha+\vec\beta)\cdot\vec\theta} \textrm{ if }
(\vec\alpha+\vec\beta)\in \Gh.\end{array}\right.  } which concludes
the proof of the lemma.\fin

Now by definition of $\Gh$, for any vector $\vec\gamma\in\Gh$, there
exists a vector $\vec\lambda_{\vec\gamma}\in\bin^{r+m}$ such that \eq{
\vec\lambda_{\vec\gamma}^T \ha{G}=\gamma } and the above property
reads: \eq{
\bra{\Psi(\vec\alpha,\tb(E\cap\overline{M}))}\rhosk\ket{\Psi(\vec\beta,\tb(E\cap\overline{M}))}
=\frac{1}{2^{\lh}}\times\left\{\begin{array}{l} 0 \textrm{ if }
(\vec\alpha+\vec\beta)\notin \Gh \\
(-1)^{\vec\lambda_{(\vec\alpha+\vec\beta)}\cdot\left(\left(\begin{subarray}{c}\vec
s\\\k\end{subarray}\right)+\ch{G}\g(E\cap M)\right)} \textrm{ if }
(\vec\alpha+\vec\beta)\in \Gh.\end{array}\right.  }

To see that the first term in (\ref{Dev}) is independent of $\k$,
recalling the definition of $\bpiw{z}$, write
\begin{eqnarray}
\lefteqn{\bra{\phiv}\bpiw{z}\rhosk\bpiw{z}\ket{\phiv}}\nonumber\\
&=&\sum_{\begin{subarray}{l}\vec\alpha,\vec\beta\in\bin^{\lh} |\\
w(\vec\alpha-\h(E\cap\overline{M}))<\dwh/2\\w(\vec\beta-\h(E\cap\overline{M}))<\dwh/2\end{subarray}}
\langle\phiv\ket{\Psi(\alpha,\tb(E\cap\overline{M}))}\bra{\Psi(\alpha,\tb(E\cap\overline{M}))}\rhosk\ket{\Psi(\beta,\tb(E\cap\overline{M}))}\times\nonumber\\
&\times& \bra{\Psi(\beta,\tb(E\cap\overline{M}))}\phiv\rangle
\end{eqnarray}
and the $\vec\alpha$'s and the $\vec\beta$'s contributing to the above
sum obey \eq{ w(\vec\alpha+\vec\beta)\leq
w(\vec\alpha-\h(E\cap\overline{M}))+w(\vec\beta-\h(E\cap\overline{M}))
<\dwh } thus $\vec\alpha+\vec\beta\notin\Gh^{*}$ (the set $\Gh^*$ has
been defined in Section \ref{privamp}). The $\vec\alpha$ and
$\vec\beta$ of the terms contributing in the sum are such that their
sum is in $\Gh$ (according to the previous lemma) but not in
$\Gh^{*}$. Since (by definition of $\Gh^{*}$) for
$\vec\alpha+\vec\beta\in\Gh\setminus\Gh^{*}$,
$\vec\lambda_{\vec\alpha+\vec\beta}$ is of the form
$\left(\begin{subarray}{c} \vec z\\\vec 0\end{subarray}\right)$ where
$\vec z\in\bin^r$, the terms
\begin{eqnarray}
\lefteqn{\langle\phiv\ket{\Psi(\alpha,\tb(E\cap\overline{M})}\bra{\Psi(\alpha,\tb(E\cap\overline{M})}\rhosk\ket{\Psi(\beta,\tb(E\cap\overline{M}))}\bra{\Psi(\beta,\tb(E\cap\overline{M}))}\phiv\rangle}\nonumber\\
&=&\frac{1}{2^{\lh}}(-1)^{\vec\lambda_{\vec\alpha+\vec\beta}\cdot\left(\left(\begin{subarray}{c}
\s\\\k\end{subarray}\right)+\ch{G}\g(E\cap
M)\right)}\langle\phiv\state{\vec\alpha,\tb(E\cap\overline{M})}\bra{\Psi(\vec\beta,\tb(E\cap\overline{M}))}{\phiv}\rangle
\end{eqnarray}
contributing in the above sum (i.e. for
$\vec\alpha+\vec\beta\in\Gh\setminus\Gh^*$) does not depend on
$\k$. Therefore $\bra{\phiv}\bpiw{z}\rhosk\bpiw{z}\ket{\phiv}$ does
not depend on $\k$. Now $\Pmpr$ is independent of $\k$ since the $m$
rows of $\ha{K}$ are linearly independent between themselves and
linearly independent of the rows of $\ha{F}$ (since $\dwh>0$ on $\R$
by definition (Eq.~(\ref{DefR}))).

Therefore, the term in the rhs.~of (\ref{epiv}) is independent of
$\k$. This concludes the proof of the property.\fin

\paragraph{The deviation from the key-independent term \newline}

We now derive an upper bound on $|\pr{\k v}(\k,v)-\pi_v|$.

\begin{property}\label{kdep}
For any $v\in\R\cap\L$ and $\k\in\bin^m$, define $\eta_{v,\k}$ as \eq{
\eta_{v,\k} \stackrel{Def}{=} \pr{v\k}(v,\k) - \pi_v.  } The modulus
of $\eta_{v,\k}$ is then upper bounded for any $v\in\R\cap\L$ and any
$\k\in\bin^m$ by \eq{ |\eta_{v,\k}| \leq
\frac{1}{2^m}\pr{v}(v)\left(r_{v,\k}(\piw{z}\ket{\phiv})+r_{v,\k}(\ket{\phiv})\right)\Big[2\sqrt{\sqrt{\gd}}+\sqrt{\gd}\Big].
}
\end{property}

\proof
For any $v\in\R\cap\L$ and $\k\in\bin^m$, we have from Equation (\ref{Dev}),
\begin{eqnarray}
\eta_{v,\k} &=& \pr{v\k}(v,\k) - \pi_v\\ &=& \frac{1}{2^m}
\Pmpr\Big[\bra{\phiv}\piw{z}\rhosk\ket{\phiv}+\bra{\phiv}\rhosk\piw{z}\ket{\phiv}-\nonumber\\
&-&\bra{\phiv}\piw{z}\rhosk\piw{z}\ket{\phiv}\Big]
\end{eqnarray}

Remarking that the second term in the bracket is only the complex
conjugate of the first term, we have
\begin{eqnarray}\label{f1}
|\eta_{v,\k}| &\leq& \frac{1}{2^m} \Pmpr\Big[
2\left|\bra{\phiv}\piw{z}\rhosk\ket{\phiv}\right|+\nonumber\\ &+&
\bra{\phiv}\piw{z}\rhosk\piw{z}\ket{\phiv}\Big].
\end{eqnarray}

Now, the first term in the bracket verifies
\begin{eqnarray}
&&\left|\bra{\phiv}\piw{z}\rhosk^{1/2}\rhosk^{1/2}\ket{\phiv}\right|\nonumber\\
&\leq&
\left\|\rhosk^{1/2}\piw{z}\ket{\phiv}\right\|\times\left\|\rhosk^{1/2}\ket{\phiv}\right\|\\
& & \mbox{ using the Schwartz inequality and the fact } \rhosk \mbox{
is Hermitian non negative} \nonumber\\ &=&
\sqrt{\bra{\phiv}\piw{z}\rhosk\piw{z}\ket{\phiv}}\times\sqrt{\bra{\phiv}\rhosk\ket{\phiv}}.\label{f2}
\end{eqnarray}

Now recalling the definition of $r_{v,\k}$, we have \eq{
\bra{\phiv}\rhosk\ket{\phiv} =
r_{v,\k}\big[\ket{\phiv}\big]\bra{\phiv}\rhos\ket{\phiv}
\label{f3}
} since $\bra{\phiv}\rhos\ket{\phiv}>0$, for any $v\in\R$ (Recall that
$\pr{v}(v)=\Pmpr\bra{\phiv}\rhos\ket{\phiv}$). And \eq{\label{f4}
\bra{\phiv}\piw{z}\rhosk\piw{z}\ket{\phiv} =
r_{v,\k}\big[\piw{z}\ket{\phiv}\big]\bra{\phiv}\piw{z}\rhos\piw{z}\ket{\phiv}
} (recall that if $\bra{\phiv}\piw{z}\rhos\piw{z}\ket{\phiv}=0$ then
$\bra{\phiv}\piw{z}\rhosk\piw{z}\ket{\phiv}=0$ as well).

The latter can be bounded using the small sphere property (Property
\ref{subsetL}). If $v\in \R\cap\L$,
\begin{eqnarray}
\lefteqn{\pr{\M}(\M)\Tr(E_{v|\Mc}
\piz{z}\rho_{|\Mc}\piz{z})}\nonumber\\ &=&
\Pmpr\bra{\phiv}\piz{z}\rhos\piz{z}\ket{\phiv}\\ &\leq&
\pr{v}(v)\sqrt{\gd}.
\end{eqnarray}

Now for $z\in\R$, $\frac{\dwh}{2}>\db$, thus $\textrm{Im}\,
\piw{z}\subset \textrm{Im}\, \piz{z}$ (refer to the beginning of
Section \ref{ssp}), that is $\piw{z}$ projects onto a space contained
in the space on which $\piz{z}$ projects. In other words,
$\Vect\{\state{\vec j,\tb} | d_{E\cap\overline{M}}(\vec j,\h)\geq
\dwh/2\}\subset \Vect\{\state{\vec j,\tb} | d_{E\cap\overline{M}}(\vec
j,\h)\geq \db\}$

Since $\rhos$ is Hermitian non negative, it implies that
\eq{
\bra{\phiv}\piw{z}\rhos\piw{z}\ket{\phiv}\leq
\bra{\phiv}\piz{z}\rhos\piz{z}\ket{\phiv}
}

Therefore, using Property \ref{subsetL}, we have, $\forall
\k\in\bin^m$, $\forall v\in\R\cap\L$, \eq{
\Pmpr\bra{\phiv}\piw{z}\rhos\piw{z}\ket{\phiv}\leq\pr{v}(v)\sqrt{\gd}.\label{f5}
}

Linking the results (\ref{f1},\ref{f2},\ref{f3},\ref{f4},\ref{f5})
together, we obtain
\begin{eqnarray}
\lefteqn{\forall \k\in\bin^m, \forall v\in\R\cap\L,}\nonumber\\
|\eta_{v,\k}| &\leq& \frac{1}{2^m}\Pmpr\Big[
2\sqrt{\frac{\pr{v}(v)}{\Pmpr}\sqrt{\gd}
r_{v,\k}\big(\piw{z}\ket{\phiv}\big)}\times\nonumber\\
&\times&\sqrt{r_{v,\k}\big(\ket{\phiv}\big)\bra{\phiv}\rhos\ket{\phiv}}+\nonumber\\
&+& \frac{\pr{v}(v)}{\Pmpr}\sqrt{\gd}r_{v,\k}(\piw{z}\ket{\phiv})\Big]
\end{eqnarray}
and using $\bra{\phiv}\rhos\ket{\phiv}=\pr{v}(v)/\Pmpr$, we get
\begin{eqnarray}
|\eta_{v,\k}| &\leq& \frac{1}{2^m}\Big[ 2\sqrt{\sqrt{\gd}}\times
\sqrt{r_{v,\k}\big(\piw{z}\ket{\phiv}\big)
r_{v,\k}\big(\ket{\phiv}\big)}+\nonumber\\
&+&\sqrt{\gd}r_{v,\k}\big(\piw{z}\ket{\phiv}\big)\Big]\pr{v}(v)\\
&\leq&
\frac{1}{2^m}\max\Big(\big\{r_{v,\k}\big(\piw{z}\ket{\phiv}\big),r_{v,\k}\big(\ket{\phiv}\big)\big\}\Big)\times\Big[2\sqrt{\sqrt{\gd}}+\sqrt{\gd}\Big]\pr{v}(v)\\
&\leq&
\frac{1}{2^m}\Big[r_{v,\k}\big(\piw{z}\ket{\phiv}\big)+r_{v,\k}\big(\ket{\phiv}\big)\Big]\times\big[2\sqrt{\sqrt{\gd}}+\sqrt{\gd}\big]\pr{v}(v).
\end{eqnarray}

This concludes our proof.\fin

%%%%%%%%%%%%%%%%%%%%%%%%%%%%%%%%%%%%%%%%%%%%%%%%
%	5.6.4	Bound on the conditional entropy
%%%%%%%%%%%%%%%%%%%%%%%%%%%%%%%%%%%%%%%%%%%%%%%%

\subsubsection{Bound on the conditional entropy}

In this section we conclude the privacy proof by deriving from the
previous result the following property.

\begin{property}
The conditional Shannon entropy of the key $\rv{\k}$ given Eve's view
$\rv{v}$ is lower bounded by \eq{ H(\rv{\k} | \rv{v}) \geq m -
\epsilon_1(N,m) \label{entropybound} } where \eq{ \epsilon_1(N,m) = 2
\left( m+\frac{1}{\ln 2} \right) \hd + 2\sqrt{ 2 \left( m+\frac{1}{\ln
2} \right) m \hd} +
m\left(\pr{v}(\overline{\R}\cap\P)+\pr{v}(\overline{\L}\cap\P)\right)
} and \eq{ \hd = 2 \sqrt{\sqrt{\gd}} + \sqrt{\gd} \quad\mbox{ as
defined previously}.  }
\end{property}

\proof We first prove that for any strictly positive real number $q$
and for any view $v\in \R\cap\L$, there exists a set
$\K_v\subset\bin^m$ such that
\begin{itemize}
\item $|\K_v| \geq 2^m (1-\frac{1}{q})$, and
\item $\forall \k\in\K_v$,
\eq{
\left| \pc{\k}{\rv{v}=v}(\k) -\frac{1}{2^m}\right| \leq \frac{1}{2^m}(2q+2) \hd.
}
\end{itemize}

From that we prove the bound on the conditional entropy (Eqn.(\ref{entropybound})).

For any view $v\in\R\cap\L$, summing over $\k\in\bin^m$ the joint
probability $\pr{\k v}(\k,v) = \pi_v+\eta_{v,\k}$, we get, using
Property \ref{kindep} \eq{ \forall v\in\R\cap\L,\quad
\sum_{\k\in\bin^m} \pr{\k
v}(\k,v)=\pr{v}(v)=2^m\pi_v+\sum_{\k\in\bin^m}\eta_{v,\k}.  }

but
\begin{eqnarray}
\left|\sum_{\k}\eta_{v,\k}\right|&\leq&\sum_{\k}|\eta_{v,\k}|\\ &\leq&
\frac{1}{2^m}\pr{v}(v)\hd\left(\sum_{\k}r_{v,\k}\left(\piw{z}\ket{\phiv}\right)+\sum_{\k}r_{v,\k}\left(\ket{\phiv}\right)\right)\\
&\leq& 2\pr{v}(v)\hd
\end{eqnarray}
using Property \ref{kdep} and the identity (\ref{RVK}).

Therefore,
\eq{
|\pr{v}(v)-2^m \pi_v|\leq 2\pr{v}(v)\hd
}
that is
\begin{eqnarray}
\lefteqn{|\pr{\k v}(\k,v)-\frac{1}{2^m}\pr{v}(v)|}\nonumber\\ &\leq&
|\pr{\k v}(\k,v)-\pi_v|+|\pi_v-\frac{1}{2^m}\pr{v}(v)|\\ &\leq&
\frac{1}{2^m}\pr{v}(v)\hd\Big[r_{v,\k}\left(\piw{z}\ket{\phiv}\right)+r_{v,\k}\left(\ket{\phiv}\right)+2\Big]
\end{eqnarray}
or \eq{
|\pc{\k}{\rv{v}=v}(\k)-\frac{1}{2^m}|\leq\frac{1}{2^m}\hd\Big[r_{v,\k}\left(\piw{z}\ket{\phiv}\right)+r_{v,\k}\left(\ket{\phiv}\right)+2\Big].
}

Let $a_{v,\k} = r_{v,\k}(\piw{z}\ket{\phiv}) +
r_{v,\k}(\ket{\phiv})$. Then using again identity (\ref{RVK}), we have
\eq{ \sum_{\k\in\bin^m} a_{v,\k} = 2^{m+1}.  }

Let $q$ be a strictly positive real number. Let $\rv{U}$ be a random
variable taking value in $\bin^m$ with uniform probability
distribution, i.e. $\forall \k\in\bin^m,\; \pr{U}(\k)=1/2^m$. Then
using Lemma~\ref{lemma1} for $\rv{U}$ with $\mu=2^{m+1}$, we find that
\eq{ \pr{U}(\K_v)\geq 1-\frac{1}{q} } where the set $\K_v$ is defined
by: \eq{ \K_v = \left\{ \k\in\bin^m \st a_{v,\k} <
2^{m+1}q\frac{1}{2^m} = 2q\right\}.  }

In other words,
\eq{
|\K_v| \geq 2^m\left( 1-\frac{1}{q}\right).
}

Let $\I$ be the set defined by
\eq{
\I = \cup_{v\in\R\cap\L} \K_v \times \{ v\} \subset \bin^m\times\Z.
}

It follows that
\eq{
\forall (\k,v)\in \I,\quad 
\left| \pc{\k}{\rv{v}=v} -\frac{1}{2^m}\right| \leq \frac{1}{2^m}(2q+2)\hd,
} 
and
\begin{eqnarray}
\pr{\k v}(\I) &=& \sum_{v\in\R\cap\L} \pr{v}(v)\pc{\k}{\rv{v}=v}(\K_v)
\\ &=& \sum_{v\in\R\cap\L} \left[
\pr{v}(v)\sum_{\k\in\K_v}\pc{\k}{\rv{v}=v}(\k)\right] \\ &\geq&
\sum_{v\in\R\cap\L} \left[ \pr{v}(v)\sum_{\k\in\K_v}
\frac{1}{2^m}\left( 1- (2q+2)\hd\right)\right]\\ &\geq&
\left(1-\frac{1}{q}\right)\left(1-(2q+2)\hd\right) \pr{v}(\R\cap\L)\\
&\geq& \left(1-\frac{1}{q}\right)\left(1-(2q+2)\hd \right)
\left(\pr{v}(\P)-\pr{v}(\overline{\R}\cap\P)-\pr{v}(\overline{\L}\cap\P)\right)
\\ &\geq&
\pr{v}(\P)-\pr{v}(\overline{\R}\cap\P)-\pr{v}(\overline{\L}\cap\P)
-\frac{1}{q} -(2q+2)\hd.\label{beth}
\end{eqnarray}

Now,
\begin{eqnarray}
H(\rv{\k} | \rv{v}) &=& -\sum_{\k,v} \pr{\k v}(\k,v)\log_2
\pc{\k}{\rv{v}=v}(\k)\\ &=& -\sum_{\k,v\in\overline{\P}} \pr{\k
v}(\k,v)\log_2 \pc{\k}{\rv{v}=v}(\k)- \sum_{\k,v\in\P} \pr{\k
v}(\k,v)\log_2 \pc{\k}{\rv{v}=v}(\k).
\end{eqnarray}

For any $v\in\overline{\P}$ and $\k\in\bin^m$, we have
$\pc{\k}{\rv{v}=v}(\k)=1/2^m$ since Alice chooses randomly and
independently the value for $\rv{\k}$ when the validation test is not
passed. Therefore,
\begin{eqnarray}
H(\rv{\k} | \rv{v}) &=& m\pr{v}(\overline{\P}) - \sum_{\k,v\in\P}
\pr{\k v}(\k,v)\log_2 \pc{\k}{\rv{v}=v}(\k)\\ &\geq&
m\pr{v}(\overline{\P}) - \sum_{(\k,v)\in\I} \pr{\k v}(\k,v)\log_2
\pc{\k}{\rv{v}=v}(\k)
\end{eqnarray}
since for any $v$ and $\k$, $-\log_2 \pc{\k}{\rv{v}=v}(\k)$ is
nonnegative. Using the relation: \be \forall (\k,v)\in\I,\quad
\pc{\k}{\rv{v}=v}(\k) =\frac{1}{2^m} (1+\xi_{\k,v})
\end{equation}
where $\xi_{\k,v}\leq (2q+2)\hd$ for any $(\k,v)\in\I$, we get
\begin{eqnarray}
H(\rv{\k} | \rv{v}) &\geq& m\left(\pr{v}(\overline{\P})+\pr{\k
v}(\I)\right)-\sum_{(\k,v)\in\I} \pr{\k v}(\k,v)\log_2(1+\xi_{\k,v})
\\ &\geq&
m\left(1-\pr{v}(\overline{\R}\cap\P)-\pr{v}(\overline{\L}\cap\P)-\frac{1}{q}-(2q+2)\hd\right)-\nonumber\\
&-&\frac{1}{\ln 2} (2q+2)\hd\\ &=& m-\left(m+\frac{1}{\ln
2}\right)(2q+2)\hd
-\frac{m}{q}-m(\pr{v}(\overline{\R}\cap\P)+\pr{v}(\overline{\L}\cap\P))\label{guimel}
\end{eqnarray}
where we used Equation (\ref{beth}) and the inequality
$\log_2(1+x)\leq\frac{|x|}{\ln 2}$ for any $x > -1$.

The above inequality holds for any positive real number $q\geq
1$. Especially it holds for \be q =
\sqrt{\frac{m}{2\left(m+\frac{1}{\ln 2}\right)\hd}}
\end{equation}
obtained by maximising the rhs.~in Eqn.~(\ref{guimel}). We therefore
obtain the bound on the conditional Shannon entropy of the key
$\rv{\k}$ given the view $\rv{v}$ \eq{ H(\rv{\k}|\rv{v}) \geq m -
\epsilon_1(N,m) } where \eq{ \epsilon_1(N,m) = 2 \left(m+\frac{1}{\ln
2}\right)\hd+2\sqrt{ 2 \left(m+\frac{1}{\ln 2}\right)
m\hd}+m\left(\pr{v}(\overline{\R}\cap\P)+\pr{v}(\overline{\L}\cap\P)\right).
}

This concludes the proof of privacy.\fin

\bigskip

{\bf Acknowledgement } This work was partially supported by the ESF
programme on quantum information theory (QIT). HI gratefully
acknowledges support provided by the European TMR Network
ERP-4061PL95-1412. NL gratefully acknowledges support provided by the
Academy of Finland under project number 43336. We would like to thank
Artur Ekert, Patrick Hayden, Michele Mosca, Nicolas Gisin for
interesting discussions and helpful comments.

%%%%%%%%%%%%%%%%%%%%%%%%%%%%%%%%%%%%%%%%%%%%%%%%%%%%%%%%%%%%%%%%%%%%%%%%%%
%		The Bibliography					 %
%%%%%%%%%%%%%%%%%%%%%%%%%%%%%%%%%%%%%%%%%%%%%%%%%%%%%%%%%%%%%%%%%%%%%%%%%%

%%%%%%%%%%%%%%%%%%%%%%%%%%%%%%%%%%%%%%%%%%%%%%%%%%%%%%%%%%%%%%%%%%%%%%%%%%
%		Appendix						 %
%%%%%%%%%%%%%%%%%%%%%%%%%%%%%%%%%%%%%%%%%%%%%%%%%%%%%%%%%%%%%%%%%%%%%%%%%%

\appendix
\section{Appendix: Binomial Tail Inequalities}
The following properties have been used throughout this paper.
\begin{property} \label{Binom1}
Let $\alpha$ be a positive number such that $0\leq\alpha\leq
\frac{1}{2}$. Then \eq{ \sum_{0\leq i\leq\alpha n}\binom{n}{i}\leq
2^{H_1(\alpha) n} } where $H_1(\alpha)=-\alpha\log_2\alpha
-(1-\alpha)\log_2 (1-\alpha)$ is the binary entropy function.
\end{property}

\begin{property} \label{Binom2}
Let $p$, $t$ be positive number such that $0 < p\leq p+t < 1$. Then
\eq{
\sum_{(p+t)n\leq i\leq n}\binom{n}{i}p^i (1-p)^{n-i}\leq e^{-2 t^2 n}.
}
\end{property}

\begin{property} \label{Binom3}
Let $p$, $t$ be positive number such that $0 < p-t\leq p < 1$. Then
\eq{
\sum_{0\leq i\leq(p-t)n}\binom{n}{i}p^i (1-p)^{n-i}\leq e^{-2 t^2 n}.
}
\end{property}

\begin{property} \label{Binom4}
Let $A$ be a set of size $|A|$. Let $B$ be a set. Suppose each element
of $A$ is contained in $B$ with probability $p$. Let $\tau$ be a
positive number such that $0<p-\tau<p<p+\tau<1$ . Then the probability
that $B$ contains more than $(p+\tau)|A|$ elements of $A$
(i.e. $|A\cap B|\geq (p+\tau)|A|$) is bounded by \eq{ \Pr(|A\cap
B|\geq (p+\tau)|A|)\leq \exp[-2\tau^2 |A|].  }

Likewise, the probability that $B$ contains less than $(p-\tau)|A|$
elements of $A$ is bounded by \eq{ \Pr(|A\cap B|\leq (p-\tau)|A|)\leq
\exp[-2\tau^2 |A|].  }
\end{property}

\proof \cite{CT}
Suppose $0\leq p\leq p+t\leq 1$, $q=1-p$.
For any $x\geq 1$, we have
\begin{eqnarray*}
\sum_{k\leq i\leq n}\binom{n}{i}p^i q^{n-i}&\leq& \sum_{k\leq i\leq
n}\binom{n}{i}p^i q^{n-i} x^{i-k}\\ &\leq&\sum_{0\leq i\leq
n}\binom{n}{i}p^i q^{n-i}x^{i-k}\\ &=&\frac{1}{x^k}(q+p x)^n\\ &\leq&
\frac{1}{x^{(p+t)n}}(q+p x)^n
\end{eqnarray*}
where $k=\lceil (p+t)n\rceil$. The minimum of the last expression as
function of $x$ ($x\geq 1$) is reached for $\x=\frac{q(p+t)}{p(q-t)}$
and the above inequality gives
\begin{equation}\label{An1}
\sum_{k\leq i\leq n}\binom{n}{i} p^i q^{n-i}\leq
\left[\left(\frac{p}{p+t}\right)^{p+t}\left(\frac{q}{q-t}\right)^{q-t}\right]^n
\end{equation}

The inequality above reads, for $p=1/2$ (therefore $q=1/2$) and
$t=\beta-1/2$ where $\beta=1-\alpha\in[1/2,1]$, \eq{ \sum_{\beta n\leq
i\leq n}\binom{n}{i}\leq 2^{n h(\beta)}.  }

Using the identity 
\begin{equation}\label{An2}
\binom{n}{i}=\frac{n!}{(n-i)! i!}=\binom{n}{n-i}
\end{equation}
and remarking that $H_1(\alpha)=H_1(1-\beta)=H_1(\beta)$, we get
Property~\ref{Binom1}: \eq{ \forall 0\leq\alpha\leq\frac{1}{2},\quad
\sum_{0\leq i\leq\alpha n}\binom{n}{i}\leq 2^{H_1(\alpha)n}.  }

Let's write (\ref{An1}) as \eq{ \sum_{k\leq i\leq n}\binom{n}{i}p^i
q^{n-i}\leq e^{n g(t)} } where \eq{ g(t) = \ln
\left[\left(\frac{p}{p+t}\right)^{p+t}\left(\frac{q}{q-t}\right)^{q-t}\right].
}

Then $g$ is $\C^\infty$ on $[0,q[$, and applying Taylor's formula at
order 2, we get \eq{ g(t) = g(0) + t g'(0)+\int_0^t g''(u)(t-u) d u.
} It is easy to check that $g(0)=g'(0)=0$ and that
$g''(u)=-\frac{1}{(p+u)(q-u)}\leq -4$ for any $u\in ]0,q[$. Therefore
\begin{eqnarray*}
g(t) &=&\int_0^t g''(u)(t-u) du\\
&\leq& -4\int_0^t (t-u) du\\
&\leq& -2 t^2.
\end{eqnarray*}

Since the exponential function is monotonically increasing, we get
\eq{ e^{g(t)}\leq e^{-2t^2}, } therefore \eq{ \sum_{(p+t)n\leq i\leq
n}\binom{n}{i} p^i q^{n-i}\leq e^{-2 t^2 n} } which gives Property
\ref{Binom2}.

Suppose now that $0< p-t\leq p< 1$. Using the Identity (\ref{An2}), we get
\begin{eqnarray*}
\sum_{0\leq i\leq (p-t) n}\binom{n}{i}p^i q^{n-i}&=&\sum_{0\leq i\leq
(p-t) n}\binom{n}{n-i}q^{n-i}p^i\\ &=& \sum_{n-(p-t)n\leq j\leq
n}\binom{n}{j} q^j p^{n-j}\\ &=& \sum_{(q+t)n\leq j\leq n}\binom{n}{j}
q^j p^{n-j},
\end{eqnarray*}
where $0 < q\leq q+t <1$. Applying Property \ref{Binom2}, we get \eq{
\sum_{0\leq i\leq(p-t)n}\binom{n}{i}p^i(1-p)^{n-i}\leq e^{-2t^2 n} }
which concludes the proofs for the binomial tail inequalities. We now
prove Property \ref{Binom4}.

The probability that $B$ contains exactly $k$ elements of $A$, for
$0\leq k\leq|A|$, reads \eq{ \Pr(|A\cap B| = k)=\binom{|A|}{k}p^k
(1-p)^{|A|-k}.  } Therefore, the probability that $A$ contains more
than $(p+\tau)|A|$ elements of $A$ reads
\begin{eqnarray}
\Pr(|A\cap B|\geq (p+\tau)|A|) &=& \sum_{(p+\tau)n\leq k\leq |A|}
	\Pr(|A\cap B| = k) \\ &=& \sum_{(p+\tau)n\leq k\leq |A|}
	\binom{|A|}{k}p^k (1-p)^{|A|-k} \\ &\leq&\exp[-2\tau^2|A|],
\end{eqnarray}
using the binomial tail inequality (Property \ref{Binom2}). Likewise,
the probability that $A$ contains less than $(p-\tau)|A|$ elements of
$A$ reads
\begin{eqnarray}
\Pr(|A\cap B|\leq (p-\tau)|A|) &=& \sum_{0\leq k\leq (p-\tau)|A|}
	\Pr(|A\cap B| = k) \\ &=& \sum_{0\leq k\leq (p-\tau)|A|}
	\binom{|A|}{k}p^k (1-p)^{|A|-k} \\ &\leq&\exp[-2\tau^2|A|],
\end{eqnarray}
using the binomial tail inequality (Property \ref{Binom3}).
This concludes the proof.\fin

\end{document}